\documentclass[11pt]{article}
\usepackage{xcolor, graphicx}
\usepackage{subfigure}
\usepackage{amsmath, amssymb}
\usepackage{mathrsfs}
\usepackage{caption}
\usepackage[figuresright]{rotating}
\usepackage{longtable}
\usepackage{geometry}
\usepackage{amsfonts}
\usepackage{color}
\usepackage{tabularx}
\usepackage{multirow}
\usepackage{multicol}
\usepackage{array}
\usepackage{cite}
\usepackage{booktabs}
\usepackage{rotating}
\usepackage{tikz}
\usepackage{epstopdf}
\usepackage[toc,page]{appendix}
\allowdisplaybreaks[4]

\renewcommand{\theequation}{\arabic{section}.\arabic{equation}}
\renewcommand{\Re}{\operatorname{Re}}
\renewcommand{\Im}{\operatorname{Im}}
\renewcommand{\theequation}{\arabic{section}.\arabic{equation}}
\def\sI{\textrm{sI}}

\def\V={{{\bf\rm{V}}}}

\def\beq{\begin{equation}}
\def\eeq{\end{equation}}
\def\bea{\begin{eqnarray}}
\def\eea{\end{eqnarray}}
\def\ba{\begin{array}}
\def\ea{\end{array}}
\def\no{\nonumber}

\def\lt{\left}
\def\rt{\right}

\title{Exact physical quantities of the XYZ spin chain in the thermodynamic limit}

\author{Zhirong Xin${}^{a}$, Junpeng Cao$^{b,c,d,e}$ \thanks{Corresponding author. E-mail: junpengcao@iphy.ac.cn}, Wen-Li Yang$^{d,f,g,h}$\thanks{Corresponding author. E-mail: wlyang@nwu.edu.cn}, and Yupeng Wang$^{b,d}$}

\begin{document}
\date{}
\maketitle
\begin{center}
${}^a$ School of Physics and Electronic Information, Baicheng Normal University, China\\
${}^b$ Beijing National Laboratory for Condensed Matter Physics, Institute of Physics, Chinese Academy of Sciences, Beijing 100190, China\\
${}^c$ School of Physical Sciences, University of Chinese Academy of Sciences, Beijing, China\\
${}^d$ Peng Huanwu Center for Fundamental Theory, Xian 710127, China\\
${}^e$ Songshan Lake Materials Laboratory, Dongguan, Guangdong 523808, China\\
${}^f$ Institute of Modern Physics, Northwest University, Xi'an 710127, China\\
${}^g$ School of Physics, Northwest University, Xi'an, 710127, China\\
${}^h$ Shaanxi Key Laboratory for Theoretical Physics Frontiers, Xi'an 710027, China
\end{center}

\begin{abstract}

The thermodynamic limits of the XYZ spin chain with periodic or twisted boundary conditions are studied.
By using the technique of characterizing the eigenvalue of the transfer matrix by the $T-Q$ relation and by the zeros of the associated polynomial, we
obtain the constraints of the Bethe roots and the zeros for the eigenvalues.
With the help of structure of Bethe roots, we obtain the distribution patterns of zeros.
Based on them, the physical quantities such as the surface energy and excitation energy are calculated.
We find that both of them depend on the parity of sites number due to the topological long-range Neel order on the Mobius manifold in the spin space.
We also check our results with those obtaining by the density matrix renormalization group.
The method provided in this paper can be applied to study the thermodynamic properties at the thermal equilibrium state with finite temperature.

\vspace{0.5truecm}

\end{abstract}

\newpage

\section{Introduction}

The quantum spin chains have many applications in the many-body theory, magnetic ordered materials, statistical mechanics, AdS/CFT duality, and non-linear science \cite{baxter2016exactly,korepin1993quantum,vsamaj2013introduction}.
Based on the exact solution of spin chains, many believable results about the fractional excitation, spinons, quantum phase transitions, helix states, and boundary states are obtained.
Another interesting progress is that by using the separation of variables for the quantum integrable systems, the correlation functions, form factors and norm in the certain spin chains are also
calculated. It has been shown that these many-body correlations have the wonderful determinant expressions.
Recently, the quantum spin chains and their exact solutions are used to study the evolution of some entangled initial states, quench dynamics, Non-Hermitian physics, quantum information and quantum computation.
We should note that during the study of exact solutions, many famous methods treating the many-body interactions
such as Bethe ansatz, $T-Q$ relation \cite{baxter1971eight,baxter1971one,baxter1972partition,baxter1972one}, and quantum inverse scattering method \cite{takhtadzhan1979quantum, korepin1993quantum, sklyanin1978anharmonic}
are proposed.

The XYZ spin chain is the general model quantifying the anisotropic couplings, and has many important applications in the quantum magnetism.
Due to the complicated spin exchanging and the competition, the quantum states and phase diagram of XYZ model are very rich.
To calculate the exact solution of the XYZ spin chain, people usually adopts the periodic or open boundary conditions. In fact, the anti-periodic boundary condition is also an important quantization scheme.
In this case, the boundary spin along one direction at one side of the chain remains unchanged while the rest components are flipped,
then are connected to the corresponding boundary spin components at the other side.
Although the real space of the system is one dimension, the spin space is three dimensions.
Thus the system can be regarded as a Mobius strip in the spin space. The twisted bond can be smoothly moved and the system has the topological translational invariant.
We should note that the $U(1)$ symmetry of the system is broken.

The integrability of the XYZ spin chain with anti-periodic boundary condition has been proved for many decades. However, how to write out the exact solution of the XYZ spin chain
is a longstanding problem due to the $U(1)$ symmetry and the traditional methods are invalid. Until recently,
the exact solution is obtained by using the off-diagonal Bethe ansatz (ODBA) \cite{cao2013off}.
The main idea is as follows. The eigenvalue of transfer matrix is an elliptic polynomial, which must satisfy some constraints.
According to the algebraic analysis, if the number of constraints equals to the order of the polynomial, then
the eigenvalue can be determined completely. By introducing some inhomogeneous parameters and using the technique of fusion,
we indeed obtain these constraints. Then, by constructing the inhomogeneous $T-Q$ relation,
the value of the transfer matrix and the energy spectrum of the anti-periodic XYZ spin chain are obtained \cite{cao2014spin,cao2013off-xyz,wang2015off}.

The next problem is to study the thermodynamic limit, because the particle number in actual materials is proportional to the Avogadro constant.
However, the Bethe ansatz equations in the anti-periodic XYZ spin chain are inhomogeneous, and we can not take the logarithm.
The traditional thermodynamic Bethe ansatz (TBA) is invalid again.

In order to overcome this difficulty, some methods are suggested. For example, the thermodynamic limit can be studied by
using the degenerate points of model parameters where the Bethe ansatz equations are reduced to the homogeneous ones \cite{li2014thermodynamic,xin2020thermodynamic},
the finite size scaling analysis of the inhomogeneous term in the $T-Q$ relation where the related contribution can be neglected if the number of sites tends to infinity \cite{qiao2018twisted,xin2018thermodynamic,sun2019surface},
and the $t-W$ scheme where the eigenvalue of transfer
matrix are parameterized by the zeros of the corresponding polynomial \cite{qiao2020exact,qiao2021exact,le2021root}.

In this paper, we study the thermodynamic limit of the XYZ spin chain with twisted boundary condition.
By comprehensive analyzing the structure of the Bethe roots and the zeros of the polynomial, we
obtain the distribution patterns.
Based on them, we obtain the density of zeros in the thermodynamic limit and study the physical quantities
such as the twisted energy and elementary excitation.
We find that the twisted and excitation energies depend on the parity of sites number due to the quasi-long-range Neel order in the system.

These results are the foundations to study the thermodynamic properties at the thermal equilibrium state. Starting from these results, we can construct the quantum transfer matrix.
By using the contour integral, we can calculate the thermodynamic quantities such as free energy, entropy, specific heat and magnetic susceptibility.

The paper is organized as follows. In the next section, the model Hamiltonian and exact solution are introduced. In section 3, we calculate the thermodynamic limit of the system with real crossing parameter $\eta$.
In section 4, we consider the pure imaginary $\eta$ case.
The concluding remarks and discussions are provided in section 5. In the appendix, we list some useful identities of the elliptic functions.

\section{The XYZ spin chain and its exact solution}
\setcounter{equation}{0}

The model Hamiltonian considered in this paper is
\begin{eqnarray}\label{Hamiltoniant}
H=\frac{1}{2} \sum^N_{j=1} \left( J_x \sigma^x_j \sigma^x_{j+1} + J_y \sigma^y_j \sigma^y_{j+1} + J_z \sigma^z_j \sigma^z_{j+1} \right), \label{Hamil}
\end{eqnarray}
where $\sigma^{\alpha}_j (\alpha =x,y,z)$ is the Pauli matrix along the $\alpha$-direction on the $j$th site, $N$ is the number of sites, $J_{\alpha}$ is the coupling constant between two sites with
the nearest neighbor
\begin{equation}\label{Jf}
  J_x= e^{i\pi \eta} \frac{ \sigma (\eta+\frac{\tau}{2})}{\sigma(\frac{\tau}{2})},\quad J_y=e^{i\pi \eta} \frac{ \sigma (\eta+\frac{1+\tau}{2})}{\sigma(\frac{1+\tau}{2})},\quad J_z=\frac{ \sigma (\eta+\frac{1}{2})}{\sigma(\frac{1}{2})},
\end{equation}
$\sigma(u)$ is the quasi-double periodic elliptic function with the definition
\begin{eqnarray}
 \sigma(u) = \theta \left[
           \begin{array}{c}
             \frac{1}{2} \\
             \frac{1}{2} \\
           \end{array}
         \right](u,\tau), \qquad \theta \left[
           \begin{array}{c}
             a \\
             b \\
           \end{array}
         \right](u,\tau)=\sum_m e^{i \pi (m+a)^2\tau +2i \pi (m+a)(u+b)}, \label{zetaf}
\end{eqnarray}
$\sigma(u+1)=-\sigma(u)$, $\sigma(u+\tau)= - e^{-2i\pi(u+\frac{\tau}{2})} \sigma(u)$,
$\eta$ is the crossing parameter and $\tau$ is a pure imaginary with $\Im(\tau) > 0$.
The boundary conditions are achieved by
\begin{equation}\label{bc}
  \sigma^{x}_{N+1}= \sigma^{\beta}_{1} \sigma^{x}_{1}  \sigma^{\beta}_{1},\quad \sigma^{y}_{N+1}= \sigma^{\beta}_{1} \sigma^{y}_{1}  \sigma^{\beta}_{1},\quad \sigma^{z}_{N+1}= \sigma^{\beta}_{1} \sigma^{z}_{1}  \sigma^{\beta}_{1}, \quad \beta=0,x,y,z,
\end{equation}
where $\beta=0$ represents the periodic boundary condition ($\sigma^{0}$ is the $2\times 2$ identity matrix), while
$\beta=x$, $\beta=y$ and $\beta=z$ represent the twisted boundary condition obtained by twisting the $\pi$-angle around $x$-, $y$- and $z$-axis, respectively.

The $R$-matrix of the XYZ spin chain is the eight-vertex $R(u)\in {\rm End}(\mathbb{C}_1^2\otimes \mathbb{C}_2^2)$ given by
\begin{eqnarray}
R_{12}(u)=\left(\begin{array}{llll}\alpha(u)&&&\delta(u)\\&\beta(u)&\gamma(u)&\\
&\gamma(u)&\beta(u)&\\ \delta(u)&&&\alpha(u)\end{array}\right),
\label{Rm}
\end{eqnarray}
where $u$ is the spectral parameter, and the non-zero matrix elements are \cite{baxter2016exactly}
\begin{eqnarray}
\alpha(u)=
  \frac{\theta\left[\begin{array}{c} 0\\\frac{1}{2}
 \end{array}\right] (u,2\tau)
 \theta\left[\begin{array}{c} \frac{1}{2}\\[2pt]\frac{1}{2}
 \end{array}\right] (u+\eta,2\tau)}{\theta\left[\begin{array}{c} 0\\\frac{1}{2}
 \end{array}\right] (0,2\tau)
 \theta\left[\begin{array}{c} \frac{1}{2}\\[2pt]\frac{1}{2}
 \end{array}\right] (\eta,2\tau)},\;
\beta(u) = \frac{\theta\left[\begin{array}{c}
 \frac{1}{2}\\[2pt]\frac{1}{2}
 \end{array}\right] (u,2\tau)
 \theta\left[\begin{array}{c} 0\\\frac{1}{2}
 \end{array}\right] (u+\eta,2\tau)}
 {\theta\left[\begin{array}{c} 0\\\frac{1}{2}
 \end{array}\right] (0,2\tau)
 \theta\left[\begin{array}{c} \frac{1}{2}\\[2pt]\frac{1}{2}
 \end{array}\right] (\eta,2\tau)}, \no\\[6pt]
\gamma(u) =
  \frac{\theta\left[\begin{array}{c} 0\\\frac{1}{2}
 \end{array}\right] (u,2\tau)
 \theta\left[\begin{array}{c} 0\\\frac{1}{2}
 \end{array}\right] (u+\eta,2\tau)}
 {\theta\left[\begin{array}{c} 0\\\frac{1}{2}
 \end{array}\right] (0,2\tau)
 \theta\left[\begin{array}{c} 0\\\frac{1}{2}
 \end{array}\right] (\eta,2\tau)}, \;
\delta(u) = \frac{\theta\left[\begin{array}{c}
 \frac{1}{2}\\[2pt]\frac{1}{2}
 \end{array}\right] (u,2\tau)
 \theta\left[\begin{array}{c} \frac{1}{2}\\[2pt]\frac{1}{2}
 \end{array}\right] (u+\eta,2\tau)}
 {\theta\left[\begin{array}{c} 0\\\frac{1}{2}
 \end{array}\right] (0,2\tau)
 \theta\left[\begin{array}{c} 0\\\frac{1}{2}
 \end{array}\right] (\eta,2\tau)}.\label{r-func}
\end{eqnarray}
The $R$-matrix (\ref{Rm}) satisfies the quantum Yang-Baxter equation (QYBE),
\begin{eqnarray}
\hspace{-1.2truecm}R_{12}(u_1-u_2)R_{13}(u_1-u_3)R_{23}(u_2-u_3)=R_{23}(u_2-u_3)R_{13}(u_1-u_3)R_{12}(u_1-u_2),\label{QYB}
\end{eqnarray}
and possesses the $Z_2$-symmetry
\begin{equation}\label{Z2-sym}
\sigma^\alpha_1 \sigma^\alpha_2 R_{12}(u)=R_{12}(u)\sigma^\alpha_1\sigma^\alpha_2,\quad \quad \alpha=x,y,z.
\end{equation}

Starting from the $R$-matrix (\ref{Rm}), we construct the transfer matrix of the XYZ spin chain as
 \begin{eqnarray}
 t(u)=tr_0\lt\{\sigma^{\beta}_0 R_{0N}(u-\theta_N)\cdots R_{01}(u-\theta_1)\rt\},\label{trans}
 \end{eqnarray}
where $tr_0$ denotes taking trace in the auxiliary space and $\{ \left. \theta_j \right| j=1,\cdots,N \}$ are the free inhomogeneous parameters.
The QYBE (\ref{QYB}) and the $Z_2$-symmetry (\ref{Z2-sym}) lead to that the transfer matrices with different spectral parameters are mutually commutative,
i.e., $[t(u),t(v)]=0$, which guarantees the integrability of the model by treating $t(u)$ as the generating functional of the conserved quantities.
The Hamiltonian (\ref{Hamil}) is generated by the transfer matrix $t(u)$ as
\begin{eqnarray}
H=\frac{\sigma(\eta)}{\sigma'(0)}\left\{ \left.\frac{\partial \ln
t(u)}{\partial u} \right|_{u=0,\{\theta_j=0\}}-\frac{1}{2}N\zeta(\eta)\right\},\label{ham}
\end{eqnarray}
where $\sigma'(0)=\left.\frac{\partial}{\partial
u}\,\sigma(u)\right|_{u=0}$ and the function $\zeta(u)$ is defined as
\begin{eqnarray}
\zeta(u)=\frac{\partial}{\partial u} \{ \ln\sigma(u) \}.\label{thef}
\end{eqnarray}

Denoting the eigenvalue of the transfer matrix $t(u)$ as $\Lambda (u)$, we have the following functional relations\cite{cao2014spin,wang2015off}
\begin{eqnarray}
  && \Lambda(\theta_j)\Lambda(\theta_j-\eta) = e^{-i\pi(\delta_{\beta,x}+\delta_{\beta,z})} a(\theta_j)d(\theta_j-\eta),\quad j=1,\cdots,N, \label{p1}\\
  && \prod^{N}_{j=1} \Lambda(\theta_j) = c \prod^{N}_{j=1} a(\theta_j),  \label{p2}\\
  && \Lambda (u+1)=(-1)^{N} e^{-i\pi(\delta_{\beta,x}+\delta_{\beta,y})} \Lambda(u), \label{p3}\\
  && \Lambda(u+\tau)=(-1)^{N} e^{-2i\pi \{ Nu +N\frac{\eta+\tau}{2} -\sum^N_{l=1}\theta_l + \frac{\delta_{\beta,y}+\delta_{\beta,z}}{2}\}} \Lambda(u), \label{p4}
\end{eqnarray}
where the functions $a(u)=d(u+\eta)$ is given by
\begin{equation}\label{adf}
 a(u) =\prod^{N}_{l=1} \frac{\sigma(u-\theta_l +\eta)}{\sigma(\eta)},
\end{equation}
and $c$ in Eq.(\ref{p2}) is the eigenvalue of the operator $U^{\beta}=\sigma^{\beta}_1\cdots \sigma^{\beta}_N$.
It is straightforward to verify that $[t(u),U^{\beta}]=0$ and $(U^{\beta})^2=\textrm{id} $. Thus $t(u)$ and $U^{\beta}$ can be diagonalized simultaneously.
We note that $c=1$ for the periodic boundary condition with $\beta=0$ and $c=\pm 1$ for the anti-periodic one with $\beta=x,y,z$.

From the definition of $R$-matrix (\ref{Rm}) and transfer matrix $t(u)$ (\ref{trans}) and using the properties of elliptic function, we deduce
\begin{equation}
  \textrm{$\Lambda(u)$, as an entire function of $u$, is an elliptic polynomial of degree $N$. }
\end{equation}
According to the quasi-periodic properties (\ref{p3}) and (\ref{p4}), we parameterize the eigenvalue $\Lambda(u)$ as
\begin{equation}\label{Lam}
  \Lambda(u)=\Lambda_0 e^{-i\pi(u+\frac{\eta}{2})(\delta_{\beta,x}+\delta_{\beta,y}+2M_1)} \prod^{N}_{l=1} \sigma(u - z_l+\frac{\eta}{2}),\quad M_1 \in \mathbb{Z}.
\end{equation}
Meanwhile, the zeros $\{z_l|l=1,\cdots ,N  \}$ in the elliptic polynomial (\ref{Lam}) yield
\begin{equation}\label{cons}
 \sum^N_{l=1}z_l = \sum^N_{j=1}\theta_j +\frac{\tau}{2} (\delta_{\beta,x}+\delta_{\beta,y} +2M_1)+\frac{1}{2} (\delta_{\beta,y}+\delta_{\beta,z}+2M_2), \quad M_1,M_2 \in \mathbb{Z}.
\end{equation}
Therefore, for a given set of inhomogeneity parameters $\{\theta_j\}$, the $N$ zeros $\{z_l \}$ and one $\Lambda_0$ can be completely determined by the constraints (\ref{p1} )-(\ref{p2}) and (\ref{cons}).
From Eqs.(\ref{ham}) and (\ref{Lam}), we obtain the energy of the Hamiltonian (\ref{Hamil})
\begin{eqnarray}
  E =- \frac{\sigma(\eta)}{\sigma'(0)} \left\{  \sum^{N}_{l=1}  \frac{ \sigma'(z_l - \frac{\eta}{2})}{ \sigma( z_l - \frac{\eta}{2}) }  + \frac{1}{2}N \frac{\sigma'(\eta)}{\sigma(\eta)} +  i\pi (\delta_{\beta,x} + \delta_{\beta,y} + 2M_1) \right\}. \label{Evalue}
\end{eqnarray}

According to the expressions (\ref{Jf}) of coupling constants, we find that the Hamiltonian (\ref{Hamiltoniant}) is Hermitian if $\eta$ is either real or pure imaginary.
Furthermore, the Hamiltonian (\ref{Hamiltoniant}) has the quasi-periodicities $H(\eta)=H(-\eta)$, $H(\eta+2)=H(\eta)$ and $H(\eta+2\tau)=e^{-4i\pi(\eta+\tau)}H(\eta)$.
Therefore, we restrict the crossing parameter $\eta$ in the interval $\eta \in (0,1)$ if $\eta$ is real, and in the interval $\Im(\eta) \in (0,\frac{\tau}{i})$ if $\eta$ is pure imaginary.
The points $\eta=0,1,\tau$ are excluded because the model (\ref{Hamiltoniant}) degenerates into the isotropic XXX spin chain at these points.

\section{Surface energy and excitation with real $\eta$}
\label{Thermo-r}
\setcounter{equation}{0}

\subsection{Patterns of zeros of $\Lambda(u)$}

We first consider that the crossing parameter $\eta$ is real. For the clarity, we parameterize the zeros as $\{ z_l \} \equiv \{i \bar{z}_l \}$.
Another trick is that we suppose that the inhomogeneity parameters are pure imaginary, $\{\theta_j \} \equiv \{\bar{\theta}_j i\} $. Then by using the crossing
symmetry of $R$-matrix, i.e. $R_{12}(u)=-\sigma^{y}_{1} R^{t_2}_{12}(-u-\eta)\sigma^{y}_{1}$ where $t_2$ means the transposition in the space $\mathbb{C}_2$, we can prove that $t(u)^{\dag} = (-1)^{N+ \delta_{\beta,x}+ \delta_{\beta,y}+ \delta_{\beta,z}} t(-u^{*}-\eta)$.
Thus the transfer matrix $t(u)$ commutates with its Hermitian conjugate $t(u)^{\dag}$, indicating that $t(u)$ is a normal matrix. Consequently, $t(u)$ and $t(u)^{\dag}$ have the same eigenstates, leading to the following relation
\begin{eqnarray}\label{Lamr}
  \Lambda(u) = (-1)^{N+\delta_{\beta,x}+ \delta_{\beta,y}+ \delta_{\beta,z}}\Lambda^{*}(-u^{*}-\eta).
\end{eqnarray}
From Eqs.(\ref{Lam}) and (\ref{Lamr}), we deduce that for any zero $\bar{z}_l$, there must be another zero $\bar{z}_j$ satisfying
\begin{equation}\label{period-r}
 \bar{z}_j=\bar{z}^{\ast}_l + m_1 \frac{\tau}{i} +m_2 i, \quad m_1, m_2 \in \mathbb{Z}.
\end{equation}
Substituting $\{z_l=\bar{z}_l i\}$ and $\{\theta_j= \bar{\theta}_j i\}$ into Eq.(\ref{cons}), we obtain that
the zeros $\{ \bar{z}_l \}$ satisfy
\begin{equation}\label{cons-r}
  \sum^{N}_{l=1} \bar{z}_l  = \sum^{N}_{j=1} \bar{\theta}_j + \frac{\tau}{2i}(\delta_{\beta,x}+\delta_{\beta,y} +2M_1)  -  \frac{i}{2} (\delta_{\beta,y}+\delta_{\beta,z}+2M_2), \quad M_1,M_2 \in \mathbb{Z},
\end{equation}
By tuning the values of integers in Eqs.(\ref{period-r})-(\ref{cons-r}) and using the quasi-periodicity of elliptic function, we restrict the zeros to the region
$\Im(\bar{z}_l)\in [-\frac{1}{2},\frac{1}{2}]$ and $\Re(\bar{z}_l)\in [-\frac{\tau}{2i},\frac{\tau}{2i}]$ in the complex plane.
From Eqs.(\ref{period-r})-(\ref{cons-r}), we also obtain that the zeros can be classified into
(i) the real ones; (ii) the ones with fixed imaginary parts $\pm \frac{i}{2}$ where its conjugate shifted by $i$ is identical to itself; (iii) the conjugate pairs $\{\bar{z}_l, \bar{z}^{*}_l\}$.

The values of real zeros [case (i)] and the ones with fixed imaginary parts $\pm \frac{i}{2}$ [case (ii)] are determined by the constraints (\ref{p1} )-(\ref{p2}) and (\ref{cons}).
Now, we analyze the detailed forms of zeros in case (iii), which can be achieved by the distribution patterns of Bethe roots \cite{le2021root}.
It was shown that if the crossing parameter $\eta$ takes following discrete values
\begin{equation}
 \eta_{L,K} =   \frac{2L+\delta_{\beta,x} +\delta_{\beta,y}}{N-2N_1}\tau + \frac{2K+ \delta_{\beta,y} +\delta_{\beta,z} }{N-2N_1}, \quad L,K,N_1\in \mathbb{Z}, \label{eta-discrete}
\end{equation}
the function $\Lambda(u)$ can be given by the homogeneous $T-Q$ relation  \cite{cao2014spin,wang2015off,xin2020thermodynamic}
\begin{eqnarray}
  \Lambda (u) &=& e^{\{i\pi (2L+\delta_{\beta,x} +\delta_{\beta,y})u+i\phi \}} \frac{\sigma^N(u+\eta)}{\sigma^N(\eta)} \frac{Q(u-\eta)}{Q(u)}  \nonumber\\
  && + e^{-i\pi(\delta_{\beta,x} +\delta_{\beta,z})} e^{\{-i\pi (2L+\delta_{\beta,x} +\delta_{\beta,y})(u+\eta)-i\phi \}} \frac{\sigma^N(u)}{\sigma^N(\eta)} \frac{Q(u+\eta)}{Q(u)}. \label{T-Q}
\end{eqnarray}
Here $Q(u)$ is the elliptic function with the definition
\begin{equation}
  Q(u)= \prod^{N_1}_{l=1} \frac{\sigma(u-u_l)}{\sigma(\eta) },
\end{equation}
and $\{u_l\}$ are the Bethe roots. Without losing generality, we take $N_1=N$. Putting $u_l \equiv \lambda_l i -\frac{\eta}{2}$ and considering the homogeneous limit $\{\bar{\theta}_j\rightarrow 0\}$, the Bethe roots and $\phi$
in Eq.(\ref{T-Q}) satisfy the following Bethe ansatz equations (BAEs) and selection rule:
\begin{eqnarray}
   && e^{-2\pi (\delta_{\beta,x} +\delta_{\beta,y} ) \lambda_j + 2i\phi } \frac{\sigma^N(i(\lambda_j-\frac{\eta}{2} i))}{\sigma^N(i(\lambda_j+\frac{\eta}{2} i))} \nonumber\\
   &&= - e^{-i\pi(\delta_{\beta,x} +\delta_{\beta,z})} \prod_{l=1}^N \frac{\sigma(i(\lambda_j- \lambda_l -\eta i)) }{\sigma(i(\lambda_j- \lambda_l +\eta i)) }, \quad j=1,\cdots,N, \label{rBAEs}  \\
   && e^{i\phi} \prod^{N}_{j=1} \frac{\sigma(i(\lambda_j-\frac{\eta}{2}i))}{\sigma(i(\lambda_j+\frac{\eta}{2}i))} = e^{\frac{ik\pi }{N}(1+\delta_{\beta,0})}, \quad k=1,\cdots,N(1+\delta_{\beta,x}+\delta_{\beta,y}+\delta_{\beta,z} ).
\end{eqnarray}
For a complex Bethe root $\lambda_j$ with an imaginary part, we readily have $\left|  \frac{\sigma(i(\lambda_j-\frac{\eta}{2}i))}{\sigma(i(\lambda_j+\frac{\eta}{2}i))} \right|\neq 1 $.
Then in the thermodynamic limit $N\rightarrow \infty$, the left hand side of BAEs (\ref{rBAEs}) tends to infinity or zero exponentially.
To keep the equation holding, the right hand side of Eq.(\ref{rBAEs}) must tend to infinity or zero with the same order,
which gives that the Bethe roots should satisfy the string hypothesis \cite{takahashi1999thermodynamics,xin2020thermodynamic, takahashi1972one}
\begin{equation}\label{string-r}
  \lambda_{j,k} = x_{j}+(\frac{n_j+1}{2} -k)\eta i +\frac{1-\nu_j}{4}i +O(e^{-\delta N}), \quad 1\leq k \leq n_j,
\end{equation}
where $x_j$ is the position of $j$-string on the real axis, $k$ means the $k$th Bethe roots in $j$-string, $O(e^{-\delta N})$ means the finite size correction,  $n_j$ is the length of $j$-string, and $\nu_j=\pm 1$ denotes the parity of $j$-string.
The center of $j$-string is the real axis if $\nu_j=1$, while is the line with fixed imaginary party $\frac{i}{2}$ in the complex plane if $\nu_j=-1$.

Substituting the zeros $\{u=i\bar{z}_j -\frac{\eta}{2} \}$ into Eq.(\ref{T-Q}), we obtain the relations among the zeros $\{\bar{z}_j\}$ and the Bethe roots $\{\lambda_j\}$
\begin{eqnarray}
   && e^{-2\pi (\delta_{\beta,x} +\delta_{\beta,y} ) \bar{z}_j + 2i\phi } \frac{\sigma^N(i(\bar{z}_j -\frac{\eta}{2} i))}{\sigma^N(i( \bar{z}_j +\frac{\eta}{2} i))} \nonumber\\
   &&   = - e^{-i\pi(\delta_{\beta,x} +\delta_{\beta,z})} \prod_{l=1}^N \frac{\sigma(i( \bar{z}_j - \lambda_l -\eta i)) }{\sigma(i( \bar{z}_j - \lambda_l +\eta i)) }, \quad j=1,\cdots,N. \label{rBAE-zero}
\end{eqnarray}
Please note that the zeros of functions $\Lambda(u)$ and $Q(u)$ can not be equal, i.e., $\bar{z}_j \neq \lambda_j$. Therefore, from the structure (\ref{string-r}) of Bethe roots,
we obtain the distribution patterns of zeros
\begin{equation}
  \bar{z}_{j} = x_{j} \pm \frac{n_j+1}{2}\eta i +\frac{1-\nu_j}{4}i +O(e^{-\delta N}).
\end{equation}
For example, if $n_j=1$ and $\nu_j=1$, the patterns of zeros read $\bar{z}_{j}=x_{j} \pm \eta i $, while if $n_j=1$ and $\nu_j=-1$, the patterns are $\bar{z}_{j}=x_{j} \pm (\frac{1}{2} - \eta)i$
where some zeros have been shifted with $i$ to ensure that $\Im(\bar{z}_{j})\in [-\frac{1}{2},\frac{1}{2}]$.

We should note that in the thermodynamic limit $N\rightarrow \infty$, by adjusting the integers $L$ and $K$ in Eq.(\ref{eta-discrete}),
the crossing parameter $\eta_{L,K}$ can take the arbitrary continuous values. Furthermore, the real values of $\eta$ can be obtained by putting $L=0$.
That is to say, the constraint (\ref{eta-discrete}) gives a lot of degenerate points. In the thermodynamic limit,
these degenerate points can tend to the actual value infinitely.
Therefore, above distribution patterns of zeros are also valid for arbitrary crossing parameter $\eta$.

From these analyses and the numerical results at the ground and the first excited states,
we find that the patterns of zeros of the system with periodic boundary condition and those with anti-periodic ones are similar.
Then we can give the uniform expressions of the thermodynamic limit results with different boundary conditions.
Additionally, we also find that the patterns can be divided into two types, corresponding to
the intervals $\eta \in (\frac{1}{2},1)$ and $\eta \in (0,\frac{1}{2})$.
Thus we should discuss them separately.

\subsection{Surface energy and excitation with $ \eta \in (\frac{1}{2},1)$}
\label{rb}

\begin{figure}[htbp]
\centering
\subfigure[]{
\includegraphics[width=3.5cm]{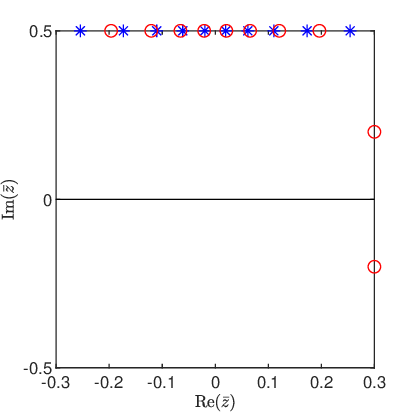}
}
\subfigure[]{
\includegraphics[width=3.5cm]{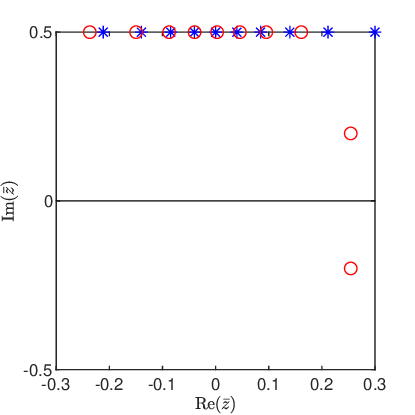}
}
\subfigure[]{
\includegraphics[width=3.5cm]{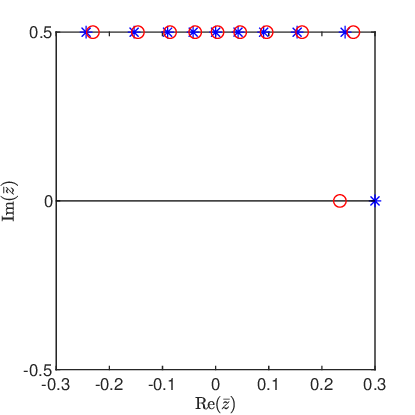}
}
\subfigure[]{
\includegraphics[width=3.5cm]{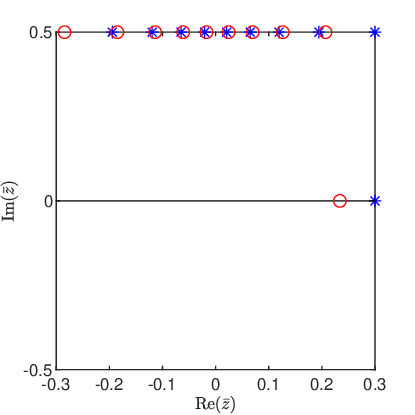}
}
\caption{
The exact numerical results of zeros $\{\bar{z}_j\}$ at the ground state (asterisks) and the first excited state (circles) with the system-size $N=10$.  The boundary conditions are (a) $\beta=0$, (b) $\beta=x$, (c) $\beta=y$, and (d) $\beta=z$.
Here $\tau=0.6i$ and $\eta=0.7$.
}\label{rb-even-fig}
\end{figure}

\begin{figure}[htbp]
\centering
\subfigure[]{
\includegraphics[width=3.5cm]{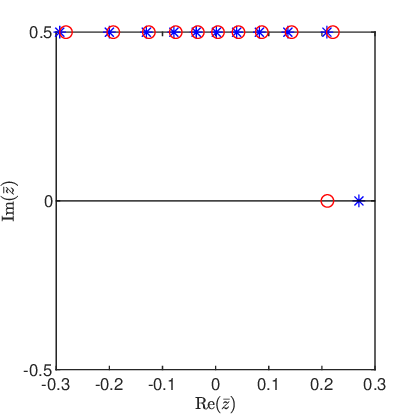}
}
\subfigure[]{
\includegraphics[width=3.5cm]{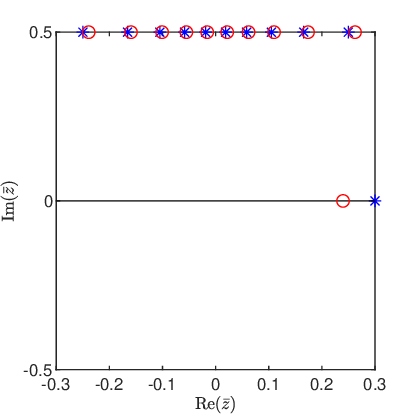}
}
\subfigure[]{
\includegraphics[width=3.5cm]{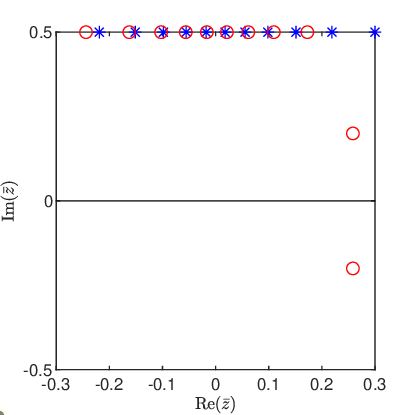}
}
\subfigure[]{
\includegraphics[width=3.5cm]{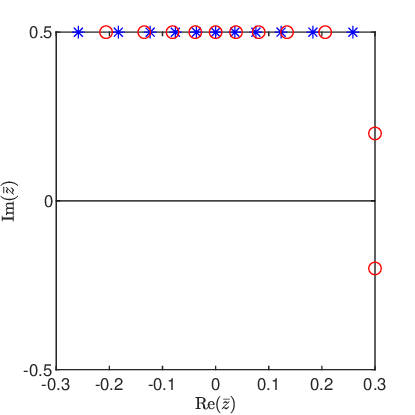}
}
\caption{
The exact numerical results of  zeros $\bar{z}_j$  at the ground state (asterisks) and the first excited state (circles) with the system-size $N=11$.  The boundary conditions are
(a) $\beta=0$, (b) $\beta=x$, (c) $\beta=y$, and (d) $\beta=z$.
Here $\tau=0.6i$ and $\eta=0.7$.
}\label{rb-odd-fig}
\end{figure}

The numerical results of the zeros $\{\bar{z}_j\}$ at the ground state and the first excited state are shown in Figs.\ref{rb-even-fig} and \ref{rb-odd-fig}.
We see that the distribution patterns of the system with periodic and
those with twisted boundaries are similar.
The main difference is the boundary strings.
The bulk strings are the zeros located on the line $\frac{i}{2}$ with the form of $\{x_l+\frac{i}{2}| l=1,\cdots,n_1 \}$ and $x_l\in [-\frac{\tau}{2i}, \frac{\tau}{2i}]$.
These zeros would be continuous in the thermodynamic limit.
Besides, there remain $n_2=N-n_1$ discrete zeros $ \{w_t| t=1,\cdots,n_2 \}$, which are real or the conjugate pairs $x\pm (\eta-\frac{1}{2})i$.
By the comprehensive analysis of the numerical results of zeros distribution and the analytical expression (\ref{cons-r}), we have
\begin{equation} \label{cons-rb}
  M_2 = -\frac{1}{2}(\delta_{\beta,y}+\delta_{\beta,z}+n_1).
\end{equation}
The discrete zeros $\{w_t\}$ satisfy the following relation
\begin{equation}\label{der}
\sum^{n_2}_{t=1} \sI(w_t +\frac{\eta}{2}i) = \sum^{n_2}_{t=1} \sI( -w_t +\frac{\eta}{2}i)= - \sum^{n_2}_{t=1} \sI(w_t-\frac{\eta}{2}i),
\end{equation}
where the function $\sI(\gamma)$ is defined by
\begin{equation}\label{sIr}
  \sI(\gamma)= \left\{ \begin{array}{cc}
                                  1, & \Im(\gamma)>0, \\
                                  0, & \Im(\gamma)=0, \\
                                  -1, & \Im(\gamma)<0.
                                \end{array} \right.
\end{equation}


Now, we are ready to calculate the physical quantities of the system in the thermodynamic limit.
From the construction of transfer matrix (\ref{trans}), we know that the eigenvalue $\Lambda(u)$ depends on two sets of
parameters. One is the set zeros and the other is the set of inhomogeneity parameters.
The inhomogeneity parameters serve as the auxiliary tools to get the thermodynamic limit results.
The physics requires that the results have the well-defined homogeneous limit.
For the real $\eta$, we choose all the inhomogeneity parameters to be pure imaginary, i.e., $\{\theta_j\}=\{\bar{\theta}_j i\}$.
Such a choice does not change the patterns of the zeros but the density of zeros \cite{qiao2021exact,le2021root}.
In the thermodynamic limit, we assume that the density of inhomogeneity parameters is $\varrho(\bar{\theta})\sim 1/N(\bar{\theta}_{j}-\bar{\theta}_{j-1})$.

Substituting the above zero patterns into Eq.(\ref{p1}), taking the logarithm, making the difference of resulted equations for $\bar{\theta}_{j}$ and $\bar{\theta}_{j-1}$, and omitting the $O(N^{-1})$ terms,
we readily have
\begin{eqnarray}
  && N\int^{\frac{\tau}{2i}}_{-\frac{\tau}{2i}} \left( A_{-\frac{1-\eta}{2}i}(u-x ) + A_{\frac{1-\eta}{2}i}(u-x )  \right) \rho(x) dx    \nonumber\\
  && + \sum^{n_2}_{t=1} \left( A_{w_t+\frac{\eta}{2}i}(u) + A_{w_t-\frac{\eta}{2}i}(u)  +  \frac{4\pi}{\tau}w_t \right) \nonumber\\
  =&&  N \int_{-\frac{\tau}{2i}}^{\frac{\tau}{2i}} \left( A_{\eta i}(u-\bar{\theta} )+ A_{-\eta i}(u-\bar{\theta} )  \right) \varrho (\bar{\theta}) d \bar{\theta}. \label{BAErb}
\end{eqnarray}
In the derivation, the constraints Eqs.(\ref{cons-r}) and (\ref{cons-rb}) have been used.
Here, $\rho(x)$ is the density of bulk strings located on the line $\frac{i}{2}$, $A_{\gamma}(x)$ is the periodic function defined as
\begin{equation}\label{Af}
  A_{\gamma}(x) = \frac{\sigma'(i(x- \gamma))}{\sigma(i(x- \gamma))}  - \frac{2\pi}{\tau}x ,
\end{equation}
and $\gamma$ is a complex number with $ \Im(\gamma) \in [-1,1]$ and $ \Re (\gamma)\in [-\frac{\tau}{2i},\frac{\tau}{2i}]$.

Taking the homogeneous limit and setting the density $\varrho (\bar{\theta})$ as the $\delta$-function, i.e., $\varrho (\bar{\theta}) \rightarrow \delta(\bar{\theta})$,
we solve the equation Eq.(\ref{BAErb}) by the via Fourier transform
and obtain the density $\rho(x)$ of the zeros as
\begin{eqnarray}
  N \tilde{\rho}(k) &=& N \frac{ \tilde{A}_{\eta i}(k) +\tilde{A}_{-\eta i}(k)}{\tilde{A}_{-\frac{1-\eta}{2}i}(k ) + \tilde{A}_{\frac{1-\eta}{2}i}(k )}  -  \frac{ \sum^{n_2}_{t=1} \left( \tilde{A}_{w_t+\frac{\eta}{2}i}(k) + \tilde{A}_{w_t-\frac{\eta}{2}i}(k) -i4\pi w_t \delta_{k,0}  \right) }{ \tilde{A}_{-\frac{1-\eta}{2}i}(k ) + \tilde{A}_{\frac{1-\eta}{2}i}(k ) }, \label{density-rb}
\end{eqnarray}
where $\tilde{\rho}(k)$ is the Fourier transformation of $\rho(x)$, and $\tilde{A}_{\gamma}(k)$ is the Fourier transformation of $A_{\gamma}(x)$
\begin{equation}
  \tilde{A}_{\gamma}(k) = \left\{ \begin{array}{cc}
                      - \pi e^{-\frac{ik\pi}{\tau}2\gamma i }  \left( \coth(\frac{ik\pi}{\tau})- \sI(\gamma)  \right) ,  & k\neq 0, \\
                     \pi( 2\gamma i+\sI(\gamma) ), & k=0,
                   \end{array}\right.
\end{equation}
and the Fourier spectrums $\{k\}$ are the integers.

Based on the pattern of zeros and using the relations (\ref{cons-r}) and (\ref{cons-rb}), the energy can be expressed by
\begin{eqnarray}
E &=&  - \frac{\sigma(\eta)}{\sigma'(0)} \left\{ N \int^{\frac{\tau}{2i}}_{-\frac{\tau}{2i}}   A_{\frac{1-\eta}{2}i}(x) \rho(x) dx  +  \sum^{n_2}_{t=1} \left( A_{-w_t-\frac{\eta}{2}i}(0)  - \frac{2\pi}{\tau} w_t  \right) \right\}   -\frac{N}{2}\frac{\sigma'(\eta)}{\sigma'(0)}     \nonumber\\
&=&  - \frac{\sigma(\eta)}{\sigma'(0)}  N \frac{i}{\tau} \sum^{\infty}_{k=- \infty}  \tilde{A}_{\frac{1-\eta}{2}i}(-k) \tilde{\rho}(k)   -\frac{N}{2}\frac{\sigma'(\eta)}{\sigma'(0)}   \nonumber\\
 && - \frac{\sigma(\eta)}{\sigma'(0)}  \sum^{n_2}_{t=1}  \frac{i}{\tau}  \left(  \sum^{\infty}_{k=- \infty} \tilde{A}_{-w_t-\frac{\eta}{2}i}(k)  + 2i\pi w_t  \right) . \label{rbE}
\end{eqnarray}
Substituting the density (\ref{density-rb}) and Eq.(\ref{der}) into (\ref{rbE}), the energy reads
\begin{eqnarray}
  E = e_{r}N + \sum^{n_2}_{t=1} E^{w}_{r}(w_t), \label{rE}
\end{eqnarray}
where $e_{r}$ is the energy density
\begin{eqnarray}
  e_{r} = - \frac{\sigma(\eta)}{\sigma'(0)} \frac{i\pi}{\tau}  \sum^{\infty}_{k= - \infty}    \tanh(\frac{ik\pi}{\tau} \eta ) \frac{  \cosh (\frac{ik\pi}{\tau}(2\eta -1)) }{\sinh(\frac{ik\pi}{\tau})}   -\frac{1}{2}\frac{\sigma'(\eta)}{\sigma'(0)}, \label{Eden-r}
\end{eqnarray}
and $E^{w}_r(w_t)$ is the energy induced by the discrete zero $w_t$
\begin{eqnarray}
  E^{w}_{r}(w_t) =   \frac{1}{2} \left( \sI(w_t+\frac{\eta}{2}i)- \sI(w_t-\frac{\eta}{2}i) \right) \frac{\sigma(\eta)}{\sigma'(0)} \frac{i \pi}{\tau}  \sum^{\infty}_{k=-\infty }    \frac{\cosh(\frac{ik\pi}{\tau}2w_ti)}{ \cosh(\frac{ik\pi}{\tau} \eta)}.\label{Eroot-r}
\end{eqnarray}
From Eq.(\ref{Eroot-r}), we see that only the discrete zeros with $\Im(w_t)\in [-\frac{\eta}{2},\frac{\eta}{2}]$ can contribute the non-zero values to the energy.

We first consider the ground state energy of the system with the boundary (\ref{bc}) where $\beta=0,x,y,z$.
From the distribution of zeros (represented by asterisks) at the ground state in Figs.\ref{rb-even-fig} and \ref{rb-odd-fig}, we see that the distribution of discrete zeros is related to the parity of $N$ and the boundary conditions.
Among them, the zero distribution for the periodic boundary condition is similar with that for the twisted boundary along the $x$-direction.
Specifically, for even $N$, there is no discrete zero, as shown by the asterisks in subgraphs (a) and (b) of Fig.\ref{rb-even-fig}.
For odd $N$, there is one real discrete zero, as shown by the asterisks in subgraphs (a) and (b) of Fig.\ref{rb-odd-fig}.

The physical explanations are as follows.
When the crossing parameter $\eta$ is real, the coupling constants (\ref{Jf}) satisfy $|J_x| > |J_y| >|J_z|$. Thus the system has the spontaneous magnetization and the easy axis is the $x$-direction.
When the couplings along $x$-direction are anti-ferromagnetic, then the directions of spins in the Neel order are antiparallel to each other. For the even $N$, two boundary spins prefer to be anti-parallel, while for the odd $N$,
two boundary spins prefer to be parallel.
We should also note that if $\beta=x$, the boundary twisted direction is the $x$-axis and the constraint (\ref{bc}) gives $\sigma_{N+1}^x=\sigma_{1}^x$, $\sigma_{N+1}^y=-\sigma_{1}^y$, $\sigma_{N+1}^z=-\sigma_{1}^z$.
We see that the dominant couplings along the $x$-direction for the periodic boundary condition and that for the twisted one are the same, due to $\sigma_{N+1}^x=\sigma_{1}^x$.
Thus the twisted bond can not change the Neel order too much.

According to Eq.(\ref{Eroot-r}), the real discrete zero should be located at the line with fixed real part $\frac{\tau}{2i}$ in the complex plane to minimize the energy in the thermodynamic limit.
Substituting the discrete zero $w_t=\frac{\tau}{2i}$ into Eq.(\ref{rE}), we obtain the ground state energy of the system with periodic boundary condition ($\beta=0$)
\begin{eqnarray}\label{Erg0x1}
E^{0}_{rg}= e_{r}N + \frac{1-(-1)^N}{2} E^{w}_{r}(\frac{\tau}{2i})+O_{0g}(N^{-1}),
\end{eqnarray}
where $O_{0g}(N^{-1})$ means the correction with the order $N^{-1}$.
The ground state energy of system with twisted boundary condition along the $x$-direction ($\beta=x$) is
\begin{eqnarray}\label{Erg0x2}
E^{x}_{rg}=e_{r}N + \frac{1-(-1)^N}{2} E^{w}_{r}(\frac{\tau}{2i})+O_{xg}(N^{-1}).
\end{eqnarray}
Comparing (\ref{Erg0x1}) and (\ref{Erg0x2}), we see that the ground state energy of the system with periodic boundary condition and that with the twisted one are the same, if
only the terms with $O(N)$ and $O(N^{0})$ orders are kept. The energy difference between $E^{0}_{rg}$ and $E^{x}_{rg}$ is in the order of $O(N^{-1})$,
which can be calculated by the integral equation (\ref{BAErb}) with the order $O(N^{-1})$.

In order to quantify the boundary effects, we define the surface energy as $E^{\beta}_{rs}=E^{\beta}_{rg}- E^{0}_{rg}$, where $\beta=x, y, z$.
From Eqs.(\ref{Erg0x1})-(\ref{Erg0x2}), we find that the surface energy with the twisted boundary along the $x$-direction is zero
\begin{equation}
  E^{x}_{rs}= 0.  \label{Ersx}
\end{equation}

The zero distribution of the system with twisted boundary condition along the $y$-direction is similar with that along the $z$-direction.
Specifically, for even $N$, there exist one real discrete zero, as shown by the asterisks in subgraphs (c) and (d) of Fig.\ref{rb-even-fig}. According to Eq.(\ref{Eroot-r}), the real discrete zero should be located at the line $\frac{\tau}{2i}$ in the thermodynamic limit to minimize the energy.
For odd $N$, there is no discrete zero, as shown by the asterisks in subgraphs (c) and (d) of Fig.\ref{rb-odd-fig}.
From Eqs.(\ref{rE})-(\ref{Eroot-r}), we obtain the ground state energies $E^{y}_{rg}$ for the twisted boundary along the $y$-direction
and $E^{z}_{rg}$ along the $z$-direction
\begin{eqnarray}
&&E^{y}_{rg}=e_{r}N + \frac{1+(-1)^N}{2} E^{w}_{r}(\frac{\tau}{2i})+O_{yg}(N^{-1}), \label{Ergyz1}\\
&&E^{z}_{rg}=e_{r}N + \frac{1+(-1)^N}{2} E^{w}_{r}(\frac{\tau}{2i})+O_{zg}(N^{-1}).\label{Ergyz2}
\end{eqnarray}
Again, the ground state energy difference between $E^{y}_{rg}$ and $E^{z}_{rg}$ is in the order $O(N^{-1})$. From Eqs.(\ref{Erg0x1}) and (\ref{Ergyz1})-(\ref{Ergyz2}),
we obtain the surface energies
\begin{eqnarray}
  E^{y}_{rs} = E^{z}_{rs} = (-1)^{N}  E^{w}_{r}(\frac{\tau}{2i}). \label{Ersyz}
\end{eqnarray}
We see that when the twisted direction is $y$- or $z$-axis, the surface energy is not the zero and $ E^{y}_{rs} = E^{z}_{rs}$.
The surface energies (\ref{Ersyz}) are different from the one (\ref{Ersx}) due to the anisotropic couplings.

Next, we consider the first excited state.
The distributions of zeros at the first excited state for the different boundary conditions are shown in Figs.\ref{rb-even-fig} and \ref{rb-odd-fig} as the circles.
From them, we see that the distributions are related to the parity of $N$ and the boundary conditions.
The distribution for the periodic boundary condition is similar with that for the twisted one along the $x$-direction. For the even $N$, there exists one conjugate pair $x\pm (\eta - \frac{1}{2})i$, which can be seen from the subgraphs (a) and (b) in Fig.\ref{rb-even-fig}.
According to Eq.(\ref{Eroot-r}), the position $x$ of the conjugate pair will be located at the line $\frac{\tau}{2i}$ in the thermodynamic limit to minimize the energy.
For the odd $N$, there exists one real discrete zero, as shown in subgraphs (a) and (b) of Fig.\ref{rb-odd-fig}.
This real discrete zero can continuously deviate from the position $\frac{\tau}{2i}$ on the real axis in the thermodynamic limit, leading to the continuous excitation.
Based on the above results, the energy at the first excited state and the excitation energy can be calculated directly.

For example, if $N$ is even, substituting the discrete zeros $x\pm (\eta - \frac{1}{2})i$ into (\ref{rE})-(\ref{Eroot-r}),
we obtain the energy $E^{0}_{re}$ at the first excited state with periodic boundary condition
and $E^{x}_{re}$ with twisted one along the $x$-direction
\begin{eqnarray}
&&E^{0}_{re}= e_{r}N + E^{w}_{r}(\frac{\tau}{2i} +(\eta - \frac{1}{2})i ) + E^{w}_{r}(\frac{\tau}{2i} -(\eta - \frac{1}{2})i )+O_{0e}(N^{-1}),\label{Ere0x1} \\
&&E^{x}_{re}=e_{r}N + E^{w}_{r}(\frac{\tau}{2i} +(\eta - \frac{1}{2})i ) + E^{w}_{r}(\frac{\tau}{2i} -(\eta - \frac{1}{2})i )+O_{xe}(N^{-1}).\label{Ere0x2}
\end{eqnarray}
Define the excitation energy as $\Delta E^{\beta}_{r} = E^{\beta}_{re} - E^{\beta}_{rg}$, where $\beta=0, x, y, z$.
From Eqs.(\ref{Erg0x1})-(\ref{Erg0x2}) and (\ref{Ere0x1})-(\ref{Ere0x2}), we obtain the excitation energy as
\begin{equation}\label{gap-rb}
 \Delta E^0_{r} = \Delta E^{x}_{r}  = E^{w}_{r}(\frac{\tau}{2i} +(\eta - \frac{1}{2})i ) + E^{w}_{r}(\frac{\tau}{2i} -(\eta - \frac{1}{2})i ).
\end{equation}
During the calculation, all the terms are kept up to the order $O(N^{-1})$.

For the odd $N$, there exists one real discrete zero at the first excited state, as shown by the circles in subgraphs (a) and (b) of Fig.\ref{rb-odd-fig}.
We denote this real zero as $w^{0}$ for the periodic boundary and $w^{x}$ for the twisted one along the $x$-direction.
According to Eq.(\ref{Eroot-r}), the related excitation mode arises from the gradual deviation of a real zero from the boundary position $\frac{\tau }{2i}$.
In the thermodynamic limit, $w^{0}$ and $w^{x}$ tend to $\frac{\tau}{2i}$ infinitely, which gives that the expressions of $E^{0}_{re}$ and $E^{x}_{re}$ can be formulated as
\begin{eqnarray}
&&E^{0}_{re}= e_{r}N + E^{w}_{r}(\frac{\tau}{2i}) + O_{0e}(N^{-1}), \label{Ere0x11} \\
&&E^{x}_{re}= e_{r}N + E^{w}_{r}(\frac{\tau}{2i}) + O_{xe}(N^{-1}).\label{Ere0x21}
\end{eqnarray}
Combining Eqs.(\ref{Erg0x1})-(\ref{Erg0x2}) and (\ref{Ere0x11})-(\ref{Ere0x21}), we derive the excitation energy as
\begin{equation}\label{nogap-rb}
\Delta E^0_{r}=0, \quad \Delta E^{x}_{r}= 0,
\end{equation}
up to the order $N^{-1}$.
Thus the excitation energy spectrum is continuous.

As explained previously, the zero distribution for the twisted boundary condition along the $y$-direction is similar with that along the $z$-direction. For the even $N$, there exists one real discrete zero, as shown in the subgraphs (c) and (d) of Fig.\ref{rb-even-fig}.
The real discrete zero can continuously deviate from the position $\frac{\tau}{2i}$ on the real axis in the thermodynamic limit, leading to the continuous excitation.
For the odd $N$, there exists one conjugate pair $x\pm (\eta - \frac{1}{2})i$, as shown in the subgraphs (c) and (d) of Fig.\ref{rb-odd-fig}.
According to Eq.(\ref{Eroot-r}), the position $x$ of the conjugate pair will be located at the line $\frac{\tau}{2i}$ in the thermodynamic limit to minimize the energy.
Based on the above results, the energy at the first excited state and the excitation energy can be calculated directly.

For example, if $N$ is even, substituting the real discrete zero $\frac{\tau}{2i}$ into Eqs.(\ref{rE})-(\ref{Eroot-r}), we obtain the energies $E^{y}_{re}$ and $E^{z}_{re}$
at the first excited state with twisted boundary condition along the $y$- and the $z$-directions, respectively,
\begin{eqnarray}
&&E^{y}_{re}= e_{r}N + E^{w}_{r}(\frac{\tau}{2i}) + O_{ye}(N^{-1}), \label{Ereyz11} \\
&&E^{z}_{re}= e_{r}N + E^{w}_{r}(\frac{\tau}{2i}) + O_{ze}(N^{-1}). \label{Ereyz21}
\end{eqnarray}
Combining Eqs.(\ref{Ergyz1})-(\ref{Ergyz2}) and (\ref{Ereyz11})-(\ref{Ereyz21}), we deduce the corresponding excitation energy as
\begin{equation}\label{nogap-rb2}
\Delta E^{y}_{r}= 0, \quad \Delta E^{z}_{r}=0.
\end{equation}
Thus the systems has the continuous energy spectrum.

If $N$ is odd, substituting the conjugate pair $x\pm (\eta - \frac{1}{2})i$ into Eqs.(\ref{rE})-(\ref{Eroot-r}),
we obtain the energy $E^{y}_{re}$ at the first excited state for the twisted boundary condition along the $y$-direction and
$E^{z}_{re}$ along the $z$-direction
\begin{eqnarray}
&&E^{y}_{re}=e_{r}N + E^{w}_{r}(\frac{\tau}{2i} +(\eta - \frac{1}{2})i ) + E^{w}_{r}(\frac{\tau}{2i} -(\eta - \frac{1}{2})i )+O_{ye}(N^{-1}), \label{Ereyz1}\\
&&E^{z}_{re}=e_{r}N + E^{w}_{r}(\frac{\tau}{2i} +(\eta - \frac{1}{2})i ) + E^{w}_{r}(\frac{\tau}{2i} -(\eta - \frac{1}{2})i )+O_{ze}(N^{-1}). \label{Ereyz2}
\end{eqnarray}
Subtracting the corresponding ground state energies, we obtain the excitation energies $\Delta E^y_{r}$, $ \Delta E^z_{r}$,
and find that both of them equal to the $\Delta E_r$ given by Eq.(\ref{gap-rb}).

\subsection{Surface energy and excitation with $ \eta \in (0,\frac{1}{2})$}
\label{rs}

\begin{figure}[htbp]
\centering
\subfigure[]{
\includegraphics[width=3.5cm]{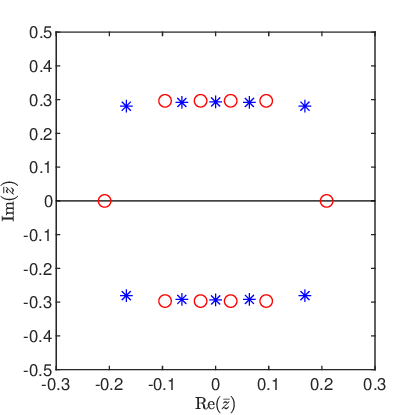}
}
\subfigure[]{
\includegraphics[width=3.5cm]{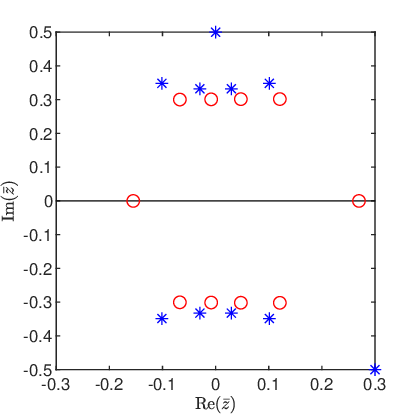}
}
\subfigure[]{
\includegraphics[width=3.5cm]{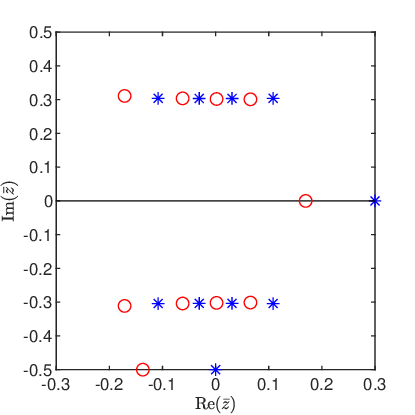}
}
\subfigure[]{
\includegraphics[width=3.5cm]{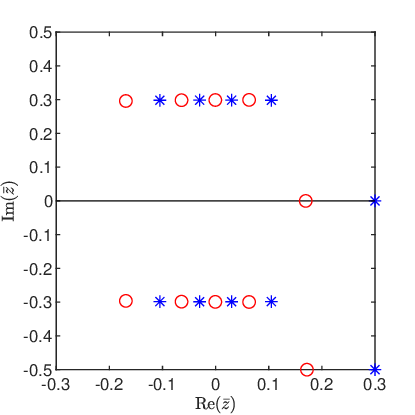}
}
\caption{ The patterns of zeros at the ground state (asterisks) and those at the first excited state (circles) for the boundary conditions (\ref{bc}) with (a) $\beta=0$, (b) $\beta=x$, (c) $\beta=y$ and (d) $\beta=z$.
Here $\tau=0.6i$, $\eta=0.3$ and $N=10$.
}\label{rs-even-fig}
\end{figure}

\begin{figure}[htbp]
\centering
\subfigure[]{
\includegraphics[width=3.5cm]{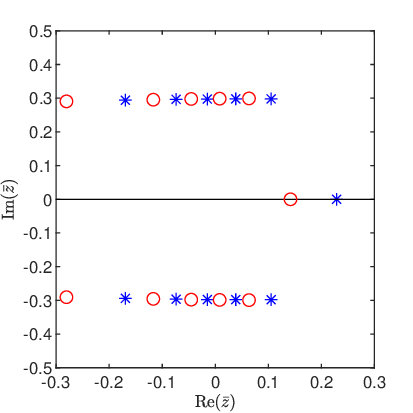}
}
\subfigure[]{
\includegraphics[width=3.5cm]{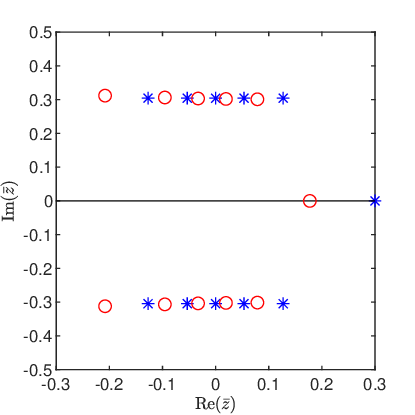}
}
\subfigure[]{
\includegraphics[width=3.5cm]{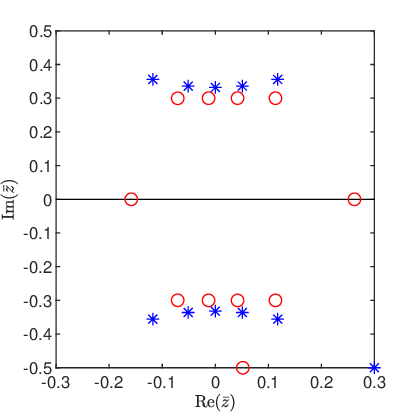}
}
\subfigure[]{
\includegraphics[width=3.5cm]{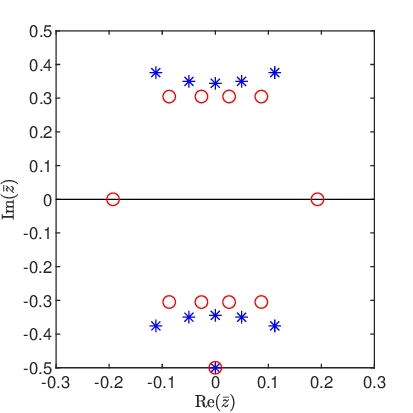}
}
\caption{
The patterns of zeros at the ground state (asterisks) and those at the first excited state (circles) for the boundary conditions (\ref{bc}) with (a) $\beta=0$, (b) $\beta=x$, (c) $\beta=y$ and (d) $\beta=z$.
Here $\tau=0.6i$, $\eta=0.3$ and $N=11$.
}\label{rs-odd-fig}
\end{figure}

Now, we consider the case of $\eta \in (0,\frac{1}{2})$. The zeros distributions at the ground state and the first excited state for the boundary conditions (\ref{bc}) with $\beta=0,x,y,z$ are shown in Figs.\ref{rs-even-fig} and \ref{rs-odd-fig}.
From them, we see that the zeros include the conjugate pairs $\{ x_l \pm \eta i| l=1,\cdots,\frac{n_1}{2}\} $ where $x_l\in [-\frac{\tau}{2i},\frac{\tau}{2i}]$ and some discrete zeros. The conjugate pairs are continuously distributed in the thermodynamic limit.
The remaining $n_2=N-n_1$ discrete zeros $ \{w_t| t=1, \cdots, n_2 \}$ are either located on the real axis or the line $\pm\frac{i}{2}$.
By the comprehensive analysis of the numerical results of zeros  distributions and the analytical expression (\ref{cons-r}), we have
\begin{eqnarray}
\sum^{N}_{l=1}\Im(\bar{z}_l) = \sum^{n_2}_{t=1}\Im(w_t)=-\frac{1}{2}(\delta_{\beta,y}+ \delta_{\beta,z}), \quad M_2=0.
\end{eqnarray}
The discrete zeros $\{w_t\}$ satisfy the following relation
\begin{eqnarray}
  && \sum^{n_2}_{t=1} \sI(w_t - \frac{\eta}{2}i)+(\delta_{\beta,y}+ \delta_{\beta,z}) = \sum^{n_2}_{t=1} \sI(-w_t-\frac{\eta}{2}i) -(\delta_{\beta,y}+ \delta_{\beta,z}) \nonumber \\
  &=& -\sum^{n_2}_{t=1} \sI(w_t+\frac{\eta}{2}i) - (\delta_{\beta,y}+ \delta_{\beta,z}). \label{der-rs}
\end{eqnarray}

Based on the above structures of zeros and using the similar procedure introduced in subsection \ref{rb}, we have
\begin{eqnarray}
 && N \int^{\frac{\tau}{2i}}_{-\frac{\tau}{2i}} \left( A_{\frac{3\eta}{2}i}(u-x) + A_{\frac{\eta}{2}i}(u-x) +A_{-\frac{\eta}{2}i}(u-x ) + A_{-\frac{3\eta}{2}i}(u-x ) \right) \rho(x) dx \nonumber \\
  && + \sum^{n_2}_{t=1} \left( A_{w_t + \frac{\eta}{2}i}(u) + A_{w_t-\frac{\eta}{2}i}(u) +\frac{4\pi}{\tau}w_t \right) + \frac{2\pi i}{\tau} (\delta_{\beta,y} + \delta_{\beta,z}) \nonumber\\
&=& N \int_{-\frac{\tau}{2i}}^{\frac{\tau}{2i}} \left( A_{\eta i}(u-\bar{\theta} )+ A_{-\eta i}(u-\bar{\theta} )  \right) \varrho (\bar{\theta}) d \bar{\theta}, \label{BAErs}
\end{eqnarray}
where $\rho(x)$ is the density of conjugate pairs and $\varrho (\bar{\theta})$ is the density of the inhomogeneity parameters.
In the homogeneous limit, the density $\varrho (\bar{\theta})$ is set as the delta function, i.e., $\varrho (\bar{\theta}) \rightarrow \delta(\bar{\theta})$. Then the density $\rho(x)$ for the homogeneous cases can be solved via the Fourier transform
and the result is
\begin{eqnarray}
  N \tilde{\rho}(k) &=& N \frac{ \tilde{A}_{\eta i}(k) +\tilde{A}_{-\eta i}(k) }{  \tilde{A}_{\frac{3\eta}{2}i}(k) + \tilde{A}_{-\frac{3\eta}{2}i}(k ) + \tilde{A}_{\frac{\eta}{2}i}(k) +\tilde{A}_{-\frac{\eta}{2}i}(k)  } \nonumber \\
   && - \frac{ \sum^{n_2}_{t=1}  \left(  \tilde{A}_{w_t + \frac{\eta}{2}i}(k) + \tilde{A}_{w_t-\frac{\eta}{2}i}(k) -4i\pi w_t \delta_{k,0} \right) + 2\pi(\delta_{\beta,y} +\delta_{\beta,z}  )\delta_{k,0} }{  \tilde{A}_{\frac{3\eta}{2}i}(k) + \tilde{A}_{-\frac{3\eta}{2}i}(k ) + \tilde{A}_{\frac{\eta}{2}i}(k) +\tilde{A}_{-\frac{\eta}{2}i}(k)  }. \label{density-rs}
\end{eqnarray}

From the structures of zeros , we also obtain the energy (\ref{Evalue}) as
\begin{eqnarray}
  E &=&  - \frac{\sigma(\eta)}{\sigma'(0)}  N \int^{\frac{\tau}{2i}}_{- \frac{\tau}{2i}} \left( A_{-\frac{3\eta}{2}i}(x)  + A_{\frac{\eta}{2}i}(x)   \right) \rho(x) dx   -\frac{N}{2}\frac{\sigma'(\eta)}{\sigma'(0)}     \nonumber\\
  && - \frac{\sigma(\eta)}{\sigma'(0)}  \left\{ \sum^{n_2}_{t=1} \left( A_{-w_t-\frac{\eta}{2}i}(0)  - \frac{2\pi}{\tau} w_t  \right) -\frac{i\pi}{\tau} (\delta_{\beta,y} +\delta_{\beta,z} )    \right\} \nonumber\\
&=&  - \frac{\sigma(\eta)}{\sigma'(0)} N \frac{i}{\tau} \sum^{\infty}_{k=-\infty} \left( \tilde{A}_{-\frac{3\eta}{2}i}(-k)  + \tilde{A}_{\frac{\eta}{2}i}(-k)   \right) \tilde{\rho}(k)   -\frac{N}{2}\frac{\sigma'(\eta)}{\sigma'(0)}   \nonumber\\
 && - \frac{\sigma(\eta)}{\sigma'(0)}  \frac{i}{\tau} \left\{ \sum^{n_2}_{t=1}  \left(  \sum^{\infty}_{k=- \infty} \tilde{A}_{-w_t-\frac{\eta}{2}i}(k)  + 2i\pi w_t  \right) -\pi (\delta_{\beta,y} +\delta_{\beta,z} )  \right\} . \label{rsE}
\end{eqnarray}
Substituting the density (\ref{density-rs}) with the constraint (\ref{der-rs}) into Eq.(\ref{rsE}), we obtain the energy $E$ which has the same forms as given by Eqs.(\ref{rE})-(\ref{Eroot-r}).
To precisely calculate the energy, the next task is to obtain the detailed forms of discrete zeros  in the region $\Im(w_t)\in [-\frac{\eta}{2},\frac{\eta}{2}]$.

We first consider the ground state energy of the system with the boundary condition (\ref{bc}) where $\beta=0, x, y, z$.
The distributions of discrete zeros are related to the parity of $N$ and the boundary conditions.
For example, the distributions for the periodic boundary condition are similar with those for the twisted boundary along the $x$-direction. This is because that the $x$-direction is the easy-axis.
Meanwhile, for the even $N$, there is no the discrete zero,
as shown by the asterisks in subgraphs (a) and (b) of Fig.\ref{rs-even-fig}. For the odd $N$, there exists one real discrete zero, as shown by the asterisks in subgraphs (a) and (b) of Fig.\ref{rs-odd-fig}.
According to Eq.(\ref{Eroot-r}), this discrete zero should be located at the line $\frac{\tau}{2i}$ in the thermodynamic limit to minimize the energy.
The ground state energies for the periodic and the twisted boundary condition along the $x$-direction can also be expressed as (\ref{Erg0x1}) and (\ref{Erg0x2}), respectively.
Therefore, the related surface energy of the twisted system is zero.

The zeros distributions of the system with twisted boundary condition along the $y$-direction and those along the $z$-direction are similar. For the even $N$, there exists one real discrete zero,
as shown by the asterisks in subgraphs (c) and (d) of Fig.\ref{rs-even-fig}. According to Eq.(\ref{Eroot-r}), the discrete zero should be located at the line $\frac{\tau}{2i}$ in the thermodynamic limit to minimize the energy.
For odd $N$, there does not exist the discrete zero, as shown by the asterisks in subgraphs (c) and (d) of Fig.\ref{rs-odd-fig}.
The corresponding ground state energies can also be expressed as Eqs.(\ref{Ergyz1})-(\ref{Ergyz2}) and the related surface energies are given by (\ref{Ersyz}).

\begin{figure}[htbp]
\centering
\subfigure[Even $N$ case]{
\includegraphics[width=6cm]{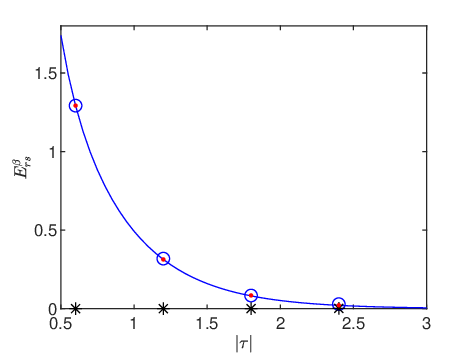}
}
\subfigure[Odd $N$ case]{
\includegraphics[width=6cm]{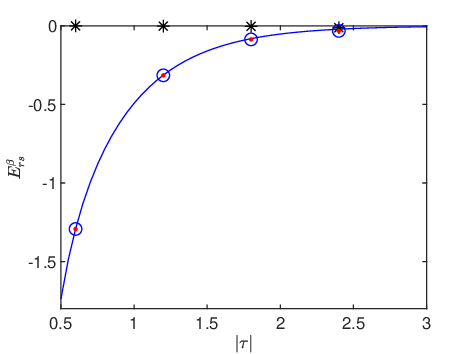}
}
\caption{
The surface energies $E^{\beta}_{rs}$ with $\beta=x, y, z$ versus the model parameter $\tau$ for $\eta=0.7$.
The blue lines denote the analytic results obtained from Eq.(\ref{Ersyz}).
The red dots, blue circles and black asterisks denote the DMRG results for $E^{y}_{rs}$, $E^{z}_{rs}$ and $E^{x}_{rs}$, respectively. (a) DMRG results with $N=150$, (b) DMRG results with $N=151$.
}\label{Esrfig}
\end{figure}

Next, we check the analytic results (\ref{Ersx}) and (\ref{Ersyz}) by using the density matrix renormalization group (DMRG) method \cite{white1992density}, and the results are shown in Fig.\ref{Esrfig}. We see that the analytic results coincide with the DMRG results
very well. Furthermore, in the trigonometric limit $\tau\rightarrow i\infty$, the coupling constants (\ref{Jf}) give $J_x=1$, $J_y=1$ and $J_z=\cos(\pi\eta)$. Thus the system reduces to the anisotropic XXZ model.
If $\eta$ is real, the related XXZ spin chain is gapless. From Fig.\ref{Esrfig}, we see that the surface energy vanishes in the trigonometric limit. Therefore, we conclude that the twisted boundary conditions
can not induce the surface energy for the XXZ model in the gapless region.

Next, we consider the first excited state. The distributions of discrete zeros are also related to the parity of $N$ and the boundary conditions, as shown in Figs.\ref{rs-even-fig} and \ref{rs-odd-fig}.
For the periodic boundary condition or the twisted one along the $x$-direction, if $N$ is even, there exist two real discrete zeros which have non-zero contributions to the energy. According to Eq.(\ref{Eroot-r}), these two discrete zeros should be $\frac{\tau}{2i}$ and $-\frac{\tau}{2i}$ in the thermodynamic limit to minimize the energy. Thus the corresponding energies $E^{\beta}_{re}$ at the first excited states obtained from Eqs.(\ref{rE})-(\ref{Eroot-r}) are
\begin{eqnarray}
&&E^{0}_{re}=e_{r}N + 2E^{w}_{r}(\frac{\tau}{2i} )+O_{0e}(N^{-1}), \label{Ere0x-small1}\\
&&E^{x}_{re}=e_{r}N + 2E^{w}_{r}(\frac{\tau}{2i} )+O_{xe}(N^{-1}).\label{Ere0x-small2}
\end{eqnarray}
Then the related excitation energies $\Delta E^{\beta}_{r}$ are
\begin{equation}\label{gap-rs}
 \Delta E^0_{r} =  \Delta E^x_{r}= 2E^{w}_{r}(\frac{\tau}{2i}).
\end{equation}
If $N$ is odd, there exist one real discrete zero which have non-zero contributions to the energy. In the thermodynamic limit, the discrete zero should be $\frac{\tau}{2i}$.
Then we obtain
\begin{eqnarray}
&&E^{0}_{re}= e_{r}N + E^{w}_{r}(\frac{\tau}{2i}) + O_{0e}(N^{-1}), \label{Ere0x12} \\
&&E^{x}_{re}= e_{r}N + E^{w}_{r}(\frac{\tau}{2i}) + O_{xe}(N^{-1}).\label{Ere0x22}
\end{eqnarray}
Thus the related excitation energies $\Delta E^0_{r}$ and $\Delta E^{x}_{r}$ are zero
\begin{equation}
\Delta E^0_{r}= 0, \quad \Delta E^{x}_{r} =0.
\end{equation}
The above analytic results agree those given in the reference \cite{xin2020thermodynamic}.

Now, we consider the twisted direction is the $y$- or $z$-axis. For the even $N$, there exists one real discrete zeros, as shown by the circles in subgraphs (c) and (d) of Fig.\ref{rs-even-fig}.
This discrete zero can continuously deviate from the position $\frac{\tau}{2i}$ on the real axis in the thermodynamic limit, leading to a continuous excitation.
For the odd $N$, there exist two real discrete zeros which contribute the non-zero values to the energy, as shown in subgraphs (c) and (d) of Fig.\ref{rs-odd-fig}. From Eq.(\ref{Eroot-r}),
these two discrete zeros should be $\frac{\tau}{2i}$ and $-\frac{\tau}{2i}$ in the thermodynamic limit to minimize the energy. The energy $E^{\beta}_{re}$ at the first excited state is
\begin{eqnarray}
&&E^{y}_{re}=e_{r}N + 2E^{w}_{r}(\frac{\tau}{2i} )+O_{ye}(N^{-1}), \label{Ereyz-small3} \\
&&E^{z}_{re}=e_{r}N + 2E^{w}_{r}(\frac{\tau}{2i} )+O_{ze}(N^{-1}). \label{Ereyz-small4}
\end{eqnarray}
We see that excitation energies $\Delta E^y_r=\Delta E^z_r$, which are also given by Eq.(\ref{gap-rs}).

\begin{figure}[!htp]
    \centering
\subfigure[]{
\includegraphics[width=6cm]{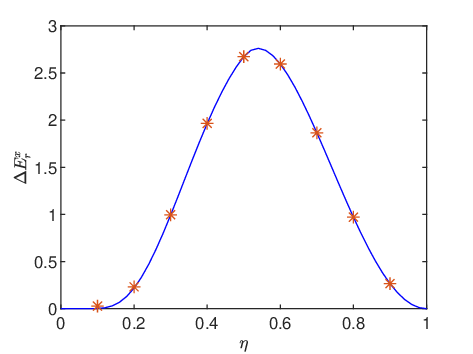}
}
\subfigure[]{
\includegraphics[width=6cm]{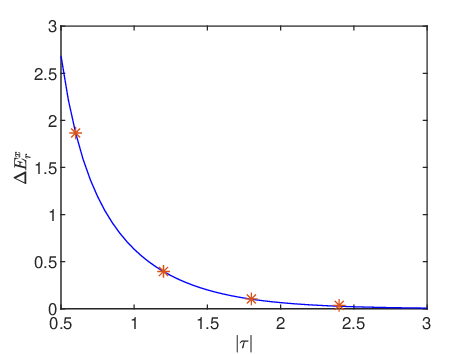}
}
\caption{
(a) The excitation energy $\Delta E^{x}_{r}$ versus the crossing parameter $\eta$ for $\tau=0.6 i$,
(b) the excitation energy $\Delta E^{x}_{r}$ versus $\tau$ for $\eta=0.7$.
The solid lines are the analytic results obtained from Eq.(\ref{gapr}) and the asterisks indicate the DMRG results with $N=310$.
}\label{fgapr}
\end{figure}

In summary, for the periodic boundary condition and the twisted one along the $x$-direction, the system has a finite excitation energy only with even $N$.
However, for the twisted boundaries along the $y$- or $z$-directions, the excitation energies are finite only when $N$ is odd. The excitation energies in these cases are equal and can be written as expressed as
\begin{equation}
  \Delta E_{r} =\left\{
              \begin{array}{cc}
              2E^{w}_{r}(\frac{\tau}{2i}  )  ,& \eta \in(0, \frac{1}{2}], \\
                E^{w}_{r}(\frac{\tau}{2i} +(\eta - \frac{1}{2})i ) + E^{w}_{r}(\frac{\tau}{2i} -(\eta - \frac{1}{2})i ) ,&  \eta \in(\frac{1}{2}, 1). \\
              \end{array}
            \right.  \label{gapr}
\end{equation}
We shall note that the results at the point of $\eta=\frac{1}{2}$ can be obtained through the analytic extension.
Now, we check the correctness of above analytic results by the DMRG method. Without losing generality, we only calculate the excitation energy $\Delta E^{x}_{r}$
and the results are shown in Fig.\ref{fgapr}. From it, we see that the analytic results (\ref{gapr}) coincide with the numerical ones very well. From Fig.\ref{fgapr}(b),
we also find that with the increase of $|\tau|$, the excitation energy tends to zero. This conclusion is consistent with the fact that
in the trigonometric limit, the related XXZ spin chain is gapless.

\section{Surface energy and excitation with pure imaginary $\eta$}
\label{Thermo-i}
\setcounter{equation}{0}

\subsection{Patterns of zeros of $\Lambda(u)$}

Assume that the inhomogeneity parameters $\{\theta_l| l=1,\cdots ,N\}$ are real. By using the properties of $R$-matrix and elliptic functions, we obtain
$t(u)^{\dag} = (-1)^{N+ \delta_{\beta,x}+ \delta_{\beta,y}+ \delta_{\beta,z}} t(u^{*}-\eta)$.
Thus, the commutation relation $[t(u),t(u)^{\dag}]=0$ holds, indicating that
\begin{equation}\label{Lami}
 \Lambda(u) = (-1)^{N+\delta_{\beta,x}+ \delta_{\beta,y}+ \delta_{\beta,z}}\Lambda^{*}(u^{*}-\eta).
\end{equation}
From the parametrization (\ref{Lam}) and using (\ref{Lami}), we find that for any zero $z_l$ of $\Lambda(u)$, there must be another zero $z_j$ satisfying
\begin{equation}\label{period-i}
z_j =z^{*}_l + m_1 \tau +m_2, \quad m_1, m_2 \in \mathbb{Z}.
\end{equation}
According to Eqs.(\ref{cons}), (\ref{period-i}) and using the quasi-periodicity of elliptic functions, we restrict the zeros in the region $\Re(z_l) \in [-\frac{1}{2},\frac{1}{2}]$ and $\Im(z_l)\in [-\frac{\tau}{2i},\frac{\tau}{2i}]$ in the complex plain.
From the numerical checking and singularity analysis, we know that the zeros $\{z_l\}$ can be classified into (i) real ones, (ii) the ones on the lines parallel to the real axis through $\pm \frac{\tau}{2}$ where their conjugates shifted by $\tau$ are identical to themselves, (iii) conjugate pairs $\{z_l,z^{\ast}_l\}$.

Now, we determine the detailed structure of the conjugate pairs. The main idea is using the degenerate points of $\eta$ and the structure of Bethe roots. At
the discrete values $\eta_{L,K}$ (\ref{eta-discrete}), the eigenvalue $\Lambda(u)$ can be expressed by a homogeneous $T-Q$ relation (\ref{T-Q}).
The singularity of the $T-Q$ relation gives the homogeneous BAEs and the associated Bethe roots can be obtained easily. Substituting
the parametrization (\ref{Lam}) into BAEs, we obtain the structure of zeros. Because there are many degenerate points.
In the thermodynamic limit $N\rightarrow\infty$, by properly adjusting the values of $L$ and taking $K=0, N_1=N$ in the constraint (\ref{eta-discrete}),
the discrete values $\eta_{L,K}$ can tend to arbitrary values of $\eta$ infinitely. Thus the results are also valid for arbitrary $\eta$.

Putting $u_l \equiv \lambda_l -\frac{\eta}{2}$ and considering the homogeneous limit $\{\theta_j \rightarrow 0 \}$, the BAEs are
\begin{eqnarray}
   && e^{ \{ 2i\pi(2L+\delta_{\beta,x} +\delta_{\beta,y}) \lambda_j +2i \phi \}} \frac{\sigma^N(\lambda_j +\frac{\eta}{2})}{\sigma^N(\lambda_j -\frac{\eta}{2})} \nonumber\\
   &&= - e^{-i\pi(\delta_{\beta,x} +\delta_{\beta,z})} \prod_{l=1}^N \frac{\sigma(\lambda_j- \lambda_l +\eta )}{\sigma( \lambda_j- \lambda_l -\eta )}, \quad j=1,\cdots,N, \label{iBAEs}  \\
   && e^{i\phi} \prod^{N}_{j=1} \frac{\sigma(\lambda_j+\frac{\eta}{2})}{\sigma(\lambda_j-\frac{\eta}{2})} = e^{\frac{ik\pi }{N}(1+\delta_{\beta,0})}, \quad k=1,\cdots,(1+\delta_{\beta,x}+\delta_{\beta,y}+\delta_{\beta,z} )N. \label{selecti}
\end{eqnarray}
The solution of BAEs gives that the Bethe roots $\{\lambda_l\}$ satisfy the string hypothesis
\begin{equation}\label{string-i}
  \lambda_{j,k} = x_{j}+(\frac{n_j+1}{2} -k)\eta  +\frac{1-\nu_j}{4}\tau  +O(e^{-\delta N}), \quad 1\leq k \leq n_j,
\end{equation}
where $x_j$ is the position of the $j$-string on the real axis, $k$ means the $k$th Bethe roots in $j$-string, $O(e^{-\delta N})$ means the finite size correction,  $n_j$ is the length of $j$-string, $\nu_j=\pm 1$ denotes the parity of $j$-string.
The center of $j$-string is the real axis if $\nu_j=1$, while is the line with fixed imaginary party $\frac{\tau}{2}$ in the complex plane if $\nu_j=-1$.

Substituting $\{u=z_j -\frac{\eta}{2} \}$ into Eq.(\ref{T-Q}), we obtain the relation between $z_j$ and $\lambda_j$
\begin{eqnarray}
   && e^{2i\pi(2L+\delta_{\beta,x} +\delta_{\beta,y}) z_j +2i \phi} \frac{\sigma^N(z_j +\frac{\eta}{2})}{\sigma^N(z_j -\frac{\eta}{2})} \nonumber \\
   &=& - e^{-i\pi(\delta_{\beta,x} +\delta_{\beta,z})} \prod_{l=1}^N \frac{\sigma(z_j- \lambda_l +\eta )}{\sigma( z_j- \lambda_l -\eta )}, \quad j=1,\cdots,N. \label{iBAE-zero}
\end{eqnarray}
We should note the zeros of functions $\Lambda(u)$ and $Q(u)$ could not be equal, i.e. $z_j \neq \lambda_j$. Thus from the structure of Bethe roots $\{ \lambda_{j,k}\}$ (\ref{string-i}), we obtain the structure of $\{z_j\}$ as
\begin{equation}
  z_{j} = x_{j} \pm \frac{n_j+1}{2}\eta  +\frac{1-\nu_j}{4}\tau +O(e^{-\delta N}).
\end{equation}
For example, if $n_j=1$ and $\nu_j=1$, the pattern of zeros read $z_{j}=x_{j} \pm \eta  $, while if $n_j=1$ and $\nu_j=-1$, the patterns are $z_{j}=x_{j} \pm (\frac{\tau}{2} - \eta) $ where some zeros have been shifted with $\tau$ to ensure that $\Im(z_j)\in[-\frac{\tau}{2i},\frac{\tau}{2i}]$.

The further analysis gives that the distributions of zeros of the eigenvalue $\Lambda(u)$
in the regions $\Im(\eta) \in (\frac{\tau}{2i},\frac{\tau}{i})$ and $\Im(\eta) \in (0,\frac{\tau}{2i})$ are different. Thus we consider them separately.

\subsection{Surface energy and excitation with $\Im(\eta) \in (\frac{\tau}{2i},\frac{\tau}{i})$}
\label{ib}

\begin{figure}[htbp]
\centering
\subfigure[]{
\includegraphics[width=3.5cm]{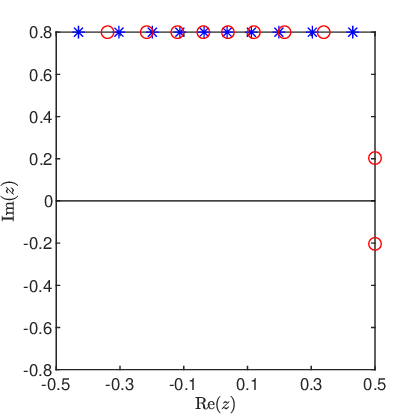}
}
\subfigure[]{
\includegraphics[width=3.5cm]{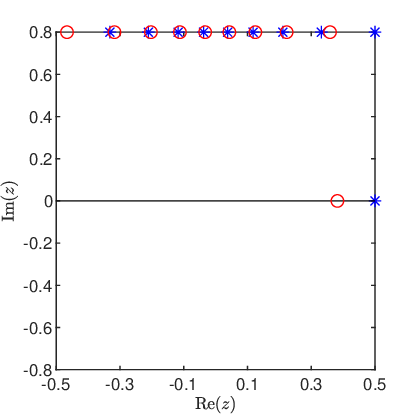}
}
\subfigure[]{
\includegraphics[width=3.5cm]{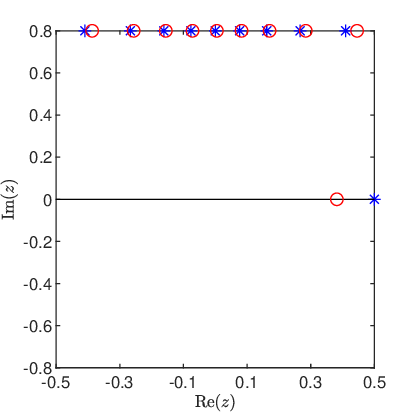}
}
\subfigure[]{
\includegraphics[width=3.5cm]{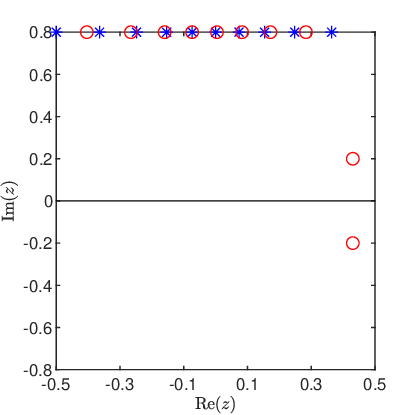}
}
\caption{
The exact numerical results of zeros at the ground state (asterisks) and the first excited state (circles) with the system-size $N=10$. The boundary conditions are (a) $\beta=0$, (b) $\beta=x$, (c) $\beta=y$ and (d) $\beta=z$.
Here $\tau=1.6i$ and $\eta=i$.
}\label{ib-even-fig}
\end{figure}

\begin{figure}[htbp]
\centering
\subfigure[]{
\includegraphics[width=3.5cm]{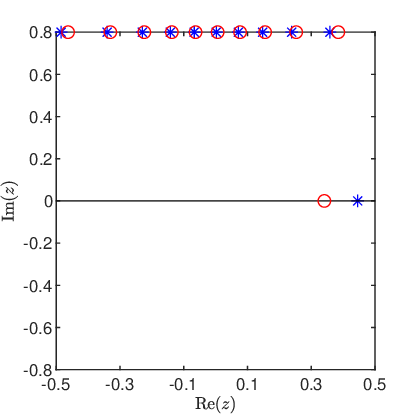}
}
\subfigure[]{
\includegraphics[width=3.5cm]{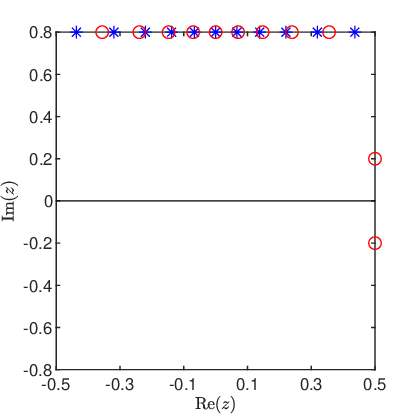}
}
\subfigure[]{
\includegraphics[width=3.5cm]{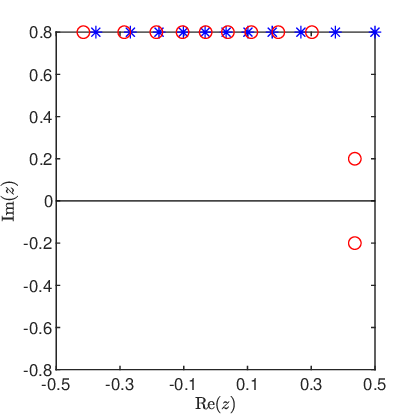}
}
\subfigure[]{
\includegraphics[width=3.5cm]{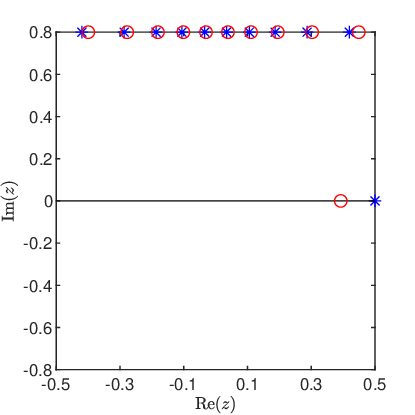}
}
\caption{
The exact numerical results of zeros at the ground state (asterisks) and the first excited state (circles) with the system-size $N=11$. The boundary conditions are (a) $\beta=0$, (b) $\beta=x$, (c) $\beta=y$ and (d) $\beta=z$.
Here $\tau=1.6i$ and $\eta=i$.
}\label{ib-odd-fig}
\end{figure}

The numerical results of the zeros of $\Lambda(u)$ at both the ground state and the first excited state with the boundary condition (\ref{bc}) for $\{\beta=0,x,y,z\}$ are shown in Figs.\ref{ib-even-fig} and \ref{ib-odd-fig}.
We see that the distribution patterns for the periodic boundary condition and those for the twisted ones are similar.
The bulk zeros are located on the line $\frac{\tau}{2}$ with the form of $\{ x_l+\frac{\tau}{2}| l=1,\cdots,n_1\}$ where $x_l\in [-\frac{1}{2}, \frac{1}{2}]$. These zeros are continuously distributed in the thermodynamic limit.
Besides, there are $n_2=N-n_1$ discrete zeros $ \{w_t| t=1,\cdots,n_2 \}$, which are real or take the forms of $x  \pm (\eta-\frac{\tau}{2})$.
Furthermore, from Eq.(\ref{cons}), we determine the values of integer $M_1$ as
\begin{equation} \label{cons-ib}
  M_1  = \frac{1}{2} (n_1- \delta_{\beta,x} - \delta_{\beta,y}).
\end{equation}
The $n_2$ discrete zeros $ \{w_t \}$ satisfy the constraints
\begin{equation}\label{der-ib}
\sum^{n_2}_{t=1} \sI(w_t+\frac{\eta}{2}) = \sum^{n_2}_{t=1} \sI(-w_t+\frac{\eta}{2})  = -\sum^{n_2}_{t=1} \sI(w_t-\frac{\eta}{2}) .
\end{equation}

Now, we assume that the inhomogeneity parameters $\{\theta_j \}$ are real. In the thermodynamic limit, the density of $\{\theta_j \}$ is determined by $\varrho(\theta)\sim 1/N(\theta_j - \theta_{j-1})$.
Based on the above structures of zeros and following the similar procedure as introduced in Section \ref{Thermo-r}, we readily have
\begin{eqnarray}
&& N  \int \left( B_{\eta}( u-\theta ) + B_{-\eta}( u-\theta ) \right) \varrho(\theta) d \theta   \nonumber\\
&=& N \int^{\frac{1}{2}}_{-\frac{1}{2}} \left( B_{ -\frac{\tau-\eta}{2} }( u-x ) +B_{\frac{\tau-\eta}{2}}( u-x ) \right) \rho(x) dx \nonumber\\
&& + \sum^{n_2}_{t=1} \left( B_{w_t+\frac{\eta}{2}}( u ) +B_{w_t-\frac{\eta}{2}}( u)  \right), \label{BAEi}
\end{eqnarray}
where $\rho(x)$ is the density of bulk zeros located on the line $ \frac{\tau}{2}$, $B_{\gamma}(x)$ is the periodic function with the definition
\begin{equation}\label{Afi}
  B_{\gamma}(x)=\frac{\sigma'(x-\gamma)}{\sigma(x-\gamma)},
\end{equation}
$\gamma$ is a complex number with $\Im(\gamma) \in [-\frac{\tau}{i} ,\frac{\tau}{i} ]$ and $\Re(\gamma) \in [-\frac{1}{2},\frac{1}{2}]$. In the derivation, Eq.(\ref{cons-ib}) has been used.

Taking the homogeneous limit $ \varrho (\theta) \rightarrow \delta(\theta)$, Eq.(\ref{BAEi}) can be solved via the Fourier transform and we obtain
\begin{equation}
 N \tilde{\rho}(k) = N \frac{\tilde{B}_{\eta}( k ) + \tilde{B}_{-\eta}( k )}{\tilde{B}_{ -\frac{\tau-\eta}{2} }( k ) +\tilde{B}_{\frac{\tau-\eta}{2}}( k )}
 - \frac{ \sum^{n_2}_{t=1} \left( \tilde{B}_{w_t+\frac{\eta}{2}}( k ) +\tilde{B}_{w_t-\frac{\eta}{2}}(k) \right) }{\tilde{B}_{ -\frac{\tau-\eta}{2} }( k ) +\tilde{B}_{\frac{\tau-\eta}{2}}( k )}, \label{density-ib}
\end{equation}
where $\tilde{B}_{\gamma}(k)$ is the Fourier transformation of $B_{\gamma}(u)$
\begin{equation}
  \tilde{B}_{\gamma}(k) = \left\{ \begin{array}{cc}
                       i\pi e^{-ik\pi 2 \gamma }  \left( \coth(ik\pi\tau)+ \sI(\gamma)  \right),   & k\neq 0, \\
                     \sI(\gamma) i\pi, & k=0,
                   \end{array}\right.
\end{equation}
and $\sI(\gamma)$ is given by Eq.(\ref{sIr}).

The patterns of zeros with the constraint (\ref{cons-ib}) also determine the energy as
\begin{eqnarray}
  E &=& - \frac{\sigma(\eta)}{\sigma'(0)}  N \int^{Q}_{-Q}  B_{-\frac{\tau-\eta}{2}}(x) \rho(x) dx   - \frac{1}{2}N \frac{\sigma'(\eta)}{\sigma'(0)}  \nonumber\\
  &&  - \frac{\sigma(\eta)}{\sigma'(0)} \left\{  \sum^{n_2}_{t=1} B_{-w_t+\frac{\eta}{2}}(0) + i\pi n_1 \right\}  \nonumber\\
  &=& - \frac{\sigma(\eta)}{\sigma'(0)}  N \sum^{\infty}_{k=- \infty}  \tilde{B}_{-\frac{\tau-\eta}{2}}(-k) \tilde{\rho}(k)   - \frac{1}{2}N \frac{\sigma'(\eta)}{\sigma'(0)}  \nonumber\\
  && - \frac{\sigma(\eta)}{\sigma'(0)} \left\{  \sum^{n_2}_{t=1} \sum^{\infty}_{k=- \infty} \tilde{B}_{-w_t+\frac{\eta}{2}}(k)  +  i\pi n_1  \right\} . \label{ibE}
\end{eqnarray}
Substituting the density (\ref{density-ib}) with the constraint (\ref{der-ib}) into (\ref{ibE}), we obtain
\begin{eqnarray}
  E = e_{i} N + \sum^{n_2}_{t=1} E^{w}_{i}(w_t), \label{iE}
\end{eqnarray}
where $e_{i}$ is the energy density
\begin{equation}\label{Eden-i}
 e_{i} =   \frac{\sigma(\eta)}{\sigma'(0)} \left\{  2 i\pi  \sum^{\infty}_{k= 1}   \tanh(ik\pi \eta)  \frac{ \cosh( ik\pi(2\eta-\tau)) }{ \sinh(ik\pi\tau) }  - \frac{1}{2} \frac{\sigma'(\eta)}{\sigma(\eta)}  \right\},
\end{equation}
and $ E_{i}^w(w_t)$ indicates the energy induced by the discrete zero $w_t$
\begin{equation}\label{Eroot-i}
E^{w}_{i} (w_t) = - \frac{1 }{2} \left( \sI(w_t+\frac{\eta}{2}) - \sI(w_t-\frac{\eta}{2}) \right)  i\pi \frac{\sigma(\eta)}{\sigma'(0)} \sum^{\infty}_{k=-\infty}  \frac{\cosh(ik\pi 2w_t)}{\cosh(ik\pi\eta)} .
\end{equation}
According to Eq.(\ref{Eroot-i}), we find that only the discrete zeros with $\Im(w_t)\in[-\frac{\eta}{2i},\frac{\eta}{2i}]$ should be taken in account when calculating the energy.

Now we consider the ground state. From the distributions of zeros (represented by asterisks) shown in Figs.\ref{ib-even-fig} and \ref{ib-odd-fig}, we see that the patterns of discrete zeros depend on the parity of $N$ and the boundary conditions.
The zero distribution for the periodic boundary condition is similar with that for the twisted boundary along the $z$-direction. Specifically, for even $N$, there is no discrete zero, as shown in subgraphs (a) and (d) of Fig.\ref{ib-even-fig}.
Neglecting the correction with the order $O(N^{-1})$, the ground state energy $E^{0}_{ig}$ for the periodic boundary condition and $E^{z}_{ig}$ for the twisted one along the $z$-direction are the same
\begin{equation}\label{Eig0z-even}
E^{0}_{ig}=E^{z}_{ig}=e_{i}N.
\end{equation}

For odd $N$, there exists one real discrete zero, as shown in subgraphs (a) and (d) of Fig.\ref{ib-odd-fig}.
From Eq.(\ref{Eroot-i}), in the thermodynamic limit, the real discrete zero should be $\frac{1}{2}$ to minimize the energy.
Therefore, if we neglect the correction with the order $O(N^{-1})$, the ground state energy with the periodic boundary condition and that with the twisted one along the $z$-direction are the same
\begin{equation}\label{Eig0z-odd}
E^{0}_{ig}=E^{z}_{ig}=e_{i}N +E^{w}_{i}(\frac{1}{2}).
\end{equation}
The energy difference between $E^{0}_{ig}$ and $E^{z}_{ig}$ can be calculated by the integral equation (\ref{BAEi}) with the order $O(N^{-1})$.
Eqs.(\ref{Eig0z-even}) and (\ref{Eig0z-odd}) give that the surface energies with the twisted boundary along the $z$-direction is zero, regardless of whether $N$ is even or odd
\begin{equation}\label{Eisz}
 E^{z}_{is} = 0.
\end{equation}
This is because that the easy-axis for real crossing parameter $\eta$ and that for imaginary $\eta$ are different.
When $\eta$ is imaginary, the coupling constant (\ref{Jf}) gives $|J_x| < |J_y| <|J_z|$.
Thus the easy-axis is the $z$-direction, and the difference between $\beta=z$ and $\beta=0$ are very small with the order $O(N^{-1})$.

The zeros patterns for the twisted boundary $\beta=x$ and that for the $\beta=y$ are similar. For the even $N$, there exists one real discrete zero, as shown by the asterisks in subgraphs (b) and (c) of Fig.\ref{ib-even-fig}. According to Eq.(\ref{Eroot-i}),
the real discrete zero should be $\frac{1}{2}$ in the thermodynamic limit to minimize the energy.
The related ground state energies are
\begin{equation}
E^{x}_{ig}= E^{y}_{ig}=e_{i}N +  E^{w}_{i}(\frac{1}{2}). \label{Eigxy-even}
\end{equation}
The energy difference between $E^{x}_{ig}$ and $E^{y}_{ig}$ is in the order $O(N^{-1})$.
From Eqs.(\ref{Eig0z-even}) and (\ref{Eigxy-even}), we obtain the surface energies
\begin{equation}
 E^{x}_{is} = E^{y}_{is} = E^{w}_{i}(\frac{1}{2}). \label{Eisxy-even}
\end{equation}
The surface energies (\ref{Eisxy-even}) are no longer the zero due to the anisotropic couplings.

For odd $N$, there are no discrete zeros, as shown by the asterisks in subgraphs (b) and (c) of Fig.\ref{ib-odd-fig}.
Neglecting the correction with the order $O(N^{-1})$, the related ground state energies are
\begin{equation}
E^{x}_{ig}= E^{y}_{ig}=e_{i}N.  \label{Eigxy-odd}
\end{equation}
From Eqs.(\ref{Eig0z-odd}) and (\ref{Eigxy-odd}), we obtain the surface energies
\begin{equation}
 E^{x}_{is} = E^{y}_{is} = -  E^{w}_{i}(\frac{1}{2}). \label{Eisxy-odd}
\end{equation}
The surface energies (\ref{Eisxy-even}) and (\ref{Eisxy-odd}) are no longer the zero due to the anisotropic couplings.

Next, we consider the first excited state.
The distributions of zeros for the different boundary conditions are shown in Figs.\ref{ib-even-fig} and \ref{ib-odd-fig} as the circles.
For the boundaries $\{\beta=0, z\}$ and even $N$, there exists one conjugate pair $x\pm (\eta - \frac{\tau}{2})$.
In the thermodynamic limit, the position $x$ of the conjugate pair will be located at the line $\frac{1}{2}$ to minimize the energy.
Substituting the discrete zeros $x\pm (\eta - \frac{\tau}{2})$ into (\ref{iE})-(\ref{Eroot-i}),
we obtain the energies $E^{\beta}_{ie}$ at the first excited state
\begin{equation}
 E^{0}_{ie} =  E^{z}_{ie} =e_{i}N + E^{w}_{i}(\frac{1}{2} +(\eta - \frac{\tau}{2}) ) + E^{w}_{i}(\frac{1}{2} -(\eta - \frac{\tau}{2}) ). \label{Eie0z1}
\end{equation}
Thus the excitation energies are
\begin{equation}\label{gap-ib}
 \Delta E^{0}_{i} =  \Delta E^{z}_{i} = E^{w}_{i}(\frac{1}{2} +(\eta - \frac{\tau}{2}) ) + E^{w}_{i}(\frac{1}{2} -(\eta - \frac{\tau}{2}) ),
\end{equation}
up to the order $O(N^{0})$.

For the boundaries $\{\beta=0, z\}$ with odd $N$, there exists one real discrete zero, which should be $\frac{1}{2}$ in the thermodynamic limit.
Consequently, the energies at the first excited state are
\begin{equation}
E^{0}_{ie}= E^{z}_{ie}=e_{i}N + E^{w}_{i}(\frac{1}{2}), \label{Eie0z11}
\end{equation}
which induce the continuous excitation
\begin{equation}\label{nogap-ib}
\Delta E^0_{i}= 0, \quad \Delta E^{z}_{i} =0.
\end{equation}
This excitation mode arises from the deviation of one real zero from the boundary.

For the boundaries $\{\beta=x, y\}$ with even $N$, there exists one real discrete zero, which tends to $\frac{1}{2}$ infinitely
in the thermodynamic limit and induces the continuous excitation, i.e., $\Delta E^x_{i}= \Delta E^{y}_{i} =0$.
While for the odd $N$, there exists one conjugate pair
$\frac{1}{2}\pm (\eta - \frac{\tau}{2})$ when $N$ tends to infinity.
Thus the energies $E^{x}_{ie}=E^{y}_{ie}$ at the first excited states can be expressed as (\ref{Eie0z1})
and the excitation energies $\Delta E^x_{i}=\Delta E^y_{i}$ can also be written as the form of Eq.(\ref{gap-ib}).

\subsection{Surface energy and excitation with $\Im(\eta) \in (0,\frac{\tau}{2i})$}
\label{is}

\begin{figure}[htbp]
\centering
\subfigure[]{
\includegraphics[width=3.5cm]{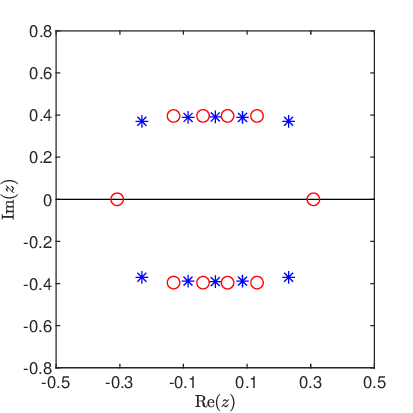}
}
\subfigure[]{
\includegraphics[width=3.5cm]{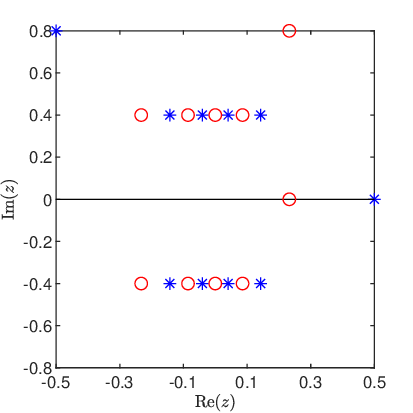}
}
\subfigure[]{
\includegraphics[width=3.5cm]{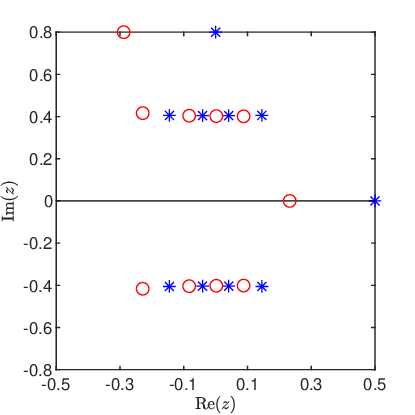}
}
\subfigure[]{
\includegraphics[width=3.5cm]{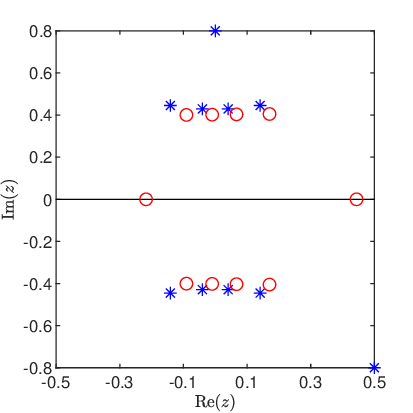}
}
\caption{
The patterns of zeros at the ground state (asterisks) and the first excited state (circles) for the boundary conditions (\ref{bc}) with (a) $\beta=0$, (b) $\beta=x$, (c) $\beta=y$ and (d) $\beta=z$.
Here $\tau=1.6i$, $\eta=0.4i$ and $N=10$.
}\label{is-even-fig}
\end{figure}

\begin{figure}[htbp]
\centering
\subfigure[]{
\includegraphics[width=3.5cm]{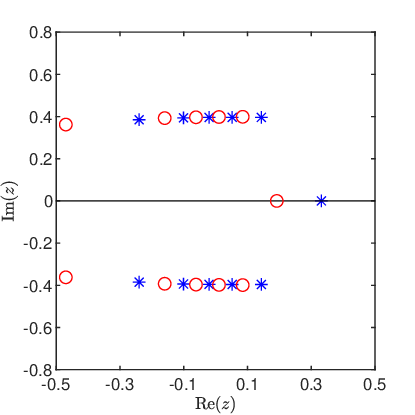}
}
\subfigure[]{
\includegraphics[width=3.5cm]{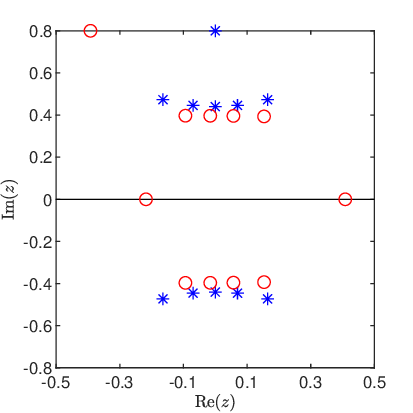}
}
\subfigure[]{
\includegraphics[width=3.5cm]{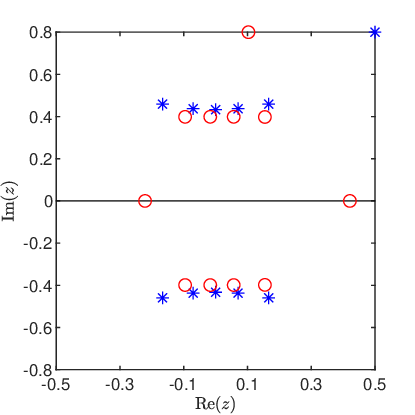}
}
\subfigure[]{
\includegraphics[width=3.5cm]{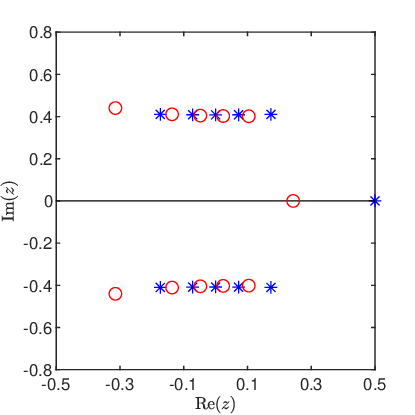}
}
\caption{
The patterns of zeros at the ground state (asterisks) and the first excited state (circles) for the boundary conditions (\ref{bc}) with (a) $\beta=0$, (b) $\beta=x$, (c) $\beta=y$ and (d) $\beta=z$.
Here $\tau=1.6i$, $\eta=0.4i$ and $N=11$.
}\label{is-odd-fig}
\end{figure}

In the region of $\Im(\eta) \in (0,\frac{\tau}{2i})$, the zeros include the conjugate pairs $ \{x_l \pm \eta| l=1, \cdots, \frac{n_1}{2} \}$ with $x_l\in [-\frac{1}{2}, \frac{1}{2}]$ and the discrete roost $ \{w_t| t=1, \cdots, n_2 \}$, which can be seen clearly from Figs.\ref{is-even-fig} and \ref{is-odd-fig}.
The conjugate pairs are continuously distributed in the thermodynamic limit, while the discrete zeros are located on the real axis or the lines $\pm\frac{\tau}{2}$.
Moreover, the zeros for different boundary conditions satisfy the constraint
\begin{equation}
\sum^{N}_{l=1} \Im(z_l)=\sum^{n_2}_{t=1} \Im(w_t)=\frac{\tau}{2i}(\delta_{\beta,x}+\delta_{\beta,y}).
\end{equation}
From Eq.(\ref{cons}), we obtain $M_1=0$. Thus the $n_2$ discrete zeros satisfy
\begin{eqnarray}
  && \sum^{n_2}_{t=1} \sI(w_t+\frac{\eta}{2})-(\delta_{\beta,x}+\delta_{\beta,y}) = \sum^{n_2}_{t=1} \sI(-w_t+\frac{\eta}{2}) +(\delta_{\beta,x}+\delta_{\beta,y}) \nonumber\\
   &=&  -\sum^{n_2}_{t=1} \sI(w_t-\frac{\eta}{2}) +(\delta_{\beta,x}+\delta_{\beta,y}). \label{der-is}
\end{eqnarray}

Based on the above structures of zeros, we obtain the integral equation for the density of zeros in the thermodynamic limit
\begin{eqnarray}
&& N  \int \left( B_{\eta}( u-\theta ) + B_{-\eta}( u-\theta ) \right) \varrho(\theta) d \theta  \nonumber\\
&=& N \int^{\frac{1}{2}}_{-\frac{1}{2}} \left\{ B_{\frac{3\eta}{2}}( u-x ) +B_{\frac{\eta}{2}}( u-x ) + B_{-\frac{\eta}{2}}( u-x ) +B_{-\frac{3\eta}{2}}( u-x ) \right\} \rho(x) dx \nonumber\\
&& + \sum^{n_2}_{t=1} \left( B_{w_t+\frac{\eta}{2}}( u) +B_{w_t -\frac{\eta}{2}}( u ) \right) -2i\pi (\delta_{\beta,x}+\delta_{\beta,y}) , \label{BAEis}
\end{eqnarray}
where $\rho(x)$ is the density of bulk conjugate pairs.
In the homogeneous limit, Eq.(\ref{BAEis}) can be solved by the Fourier transform and the solution of density is
\begin{eqnarray}
  N \tilde{\rho}(k)  &=& N \frac{\tilde{B}_{\eta}( k ) + \tilde{B}_{-\eta}( k )}{\tilde{B}_{\frac{3\eta}{2}}( k ) +\tilde{B}_{-\frac{3\eta}{2}}( k ) +\tilde{B}_{\frac{\eta}{2}}( k ) + \tilde{B}_{-\frac{\eta}{2}}( k )  } \nonumber\\
   && -    \frac{ \sum^{n_2}_{t=1}  \left( \tilde{B}_{w_t+\frac{\eta}{2}}( k ) +\tilde{B}_{w_t-\frac{\eta}{2}}( k ) \right)  -2i\pi (\delta_{\beta,x}+\delta_{\beta,y}) \delta_{k,0} }{\tilde{B}_{\frac{3\eta}{2}}( k ) +\tilde{B}_{-\frac{3\eta}{2}}( k ) +\tilde{B}_{\frac{\eta}{2}}( k ) + \tilde{B}_{-\frac{\eta}{2}}( k )  }. \label{density-is}
\end{eqnarray}
From the structures of zeros, we also obtain the energy (\ref{Evalue}) as
\begin{eqnarray}
  E &=& - \frac{\sigma(\eta)}{\sigma'(0)}  N \int^{Q}_{-Q}  \left( B_{-\frac{\eta}{2} }(x) +  B_{\frac{3\eta}{2} }(x) \right) \rho(x) dx   -  \frac{1}{2}N \frac{\sigma'(\eta)}{\sigma'(0)} \nonumber\\
  && - \frac{\sigma(\eta)}{\sigma'(0)} \left\{  \sum^{n_2}_{t=1}  B_{- w_t +\frac{\eta}{2} }(0) +i\pi (\delta_{\beta,x}+\delta_{\beta,y}) \right\}   \nonumber\\
  &=& - \frac{\sigma(\eta)}{\sigma'(0)}  N \sum^{\infty}_{k=-\infty}  \left( \tilde{B}_{-\frac{\eta}{2} }(-k) +  \tilde{B}_{\frac{3\eta}{2} }(-k) \right) \tilde{\rho}(k) -  \frac{1}{2}N \frac{\sigma'(\eta)}{\sigma'(0)}  \nonumber\\
   && - \frac{\sigma(\eta)}{\sigma'(0)}  \left\{  \sum^{n_2}_{t=1} \sum^{\infty}_{k=-\infty} \tilde{B}_{- w_t +\frac{\eta}{2} }(k) +i\pi (\delta_{\beta,x}+\delta_{\beta,y}) \right\} . \label{isE}
\end{eqnarray}
Substituting the density (\ref{density-is}) with constraint (\ref{der-is}) into (\ref{isE}), we find that the energy $E$ can still be expressed by the Eqs.(\ref{iE})-(\ref{Eroot-i}).
Similar to the discussion in subsection \ref{ib}, only the discrete zeros with $\Im(w_t)\in[-\frac{\eta}{2i},\frac{\eta}{2i}]$ should be taken in account when calculating the energy.

We first consider the ground state energy of the system with boundary conditions (\ref{bc}) where $\beta=0,x,y,z$.
The distributions of discrete zeros also depend on the different parities of $N$ and the different boundaries.
For the boundaries $\beta=0, z$ with even $N$, there is no the discrete zero satisfying $\Im(w_t)\in[-\frac{\eta}{2i},\frac{\eta}{2i}]$, as shown by the asterisks in subgraphs (a) and (d) of Fig.\ref{is-even-fig}.
Neglecting the correction with the order $O(N^{-1})$, the ground state energy $E^{0}_{ig}$ for the periodic boundary condition and $E^{z}_{ig}$ for the twisted one along the $z$-direction can also be expressed by Eq.(\ref{Eig0z-even}).

For the boundaries $\beta=0, z$ with odd $N$, there is one real discrete zero, as shown by the asterisks in subgraphs (a) and (d) of Fig.\ref{is-odd-fig}.
According to Eq.(\ref{Eroot-i}), the real discrete zero should be $\frac{1}{2}$ in the thermodynamic limit to minimize the energy.
From Eqs.(\ref{iE})-(\ref{Eroot-i}), the ground state energy $E^{0}_{ig}$ and $E^{z}_{ig}$ can also be expressed as (\ref{Eig0z-odd}).
Therefore, the related surface energy of the twisted system is zero.

For the boundaries $\beta=x, y$ with even $N$, there exists one real discrete zero, as shown by the asterisks in subgraphs (b) and (c) of Fig.\ref{is-even-fig}.
According to Eq.(\ref{Eroot-i}), the real discrete zero should be $\frac{1}{2}$ in the thermodynamic limit.
The ground state energies and surface energies are also expressed by Eqs.(\ref{Eigxy-even}) and (\ref{Eisxy-even}), respectively.

For the boundaries $\beta=x, y$ with odd $N$, there is no the discrete  zero satisfying $\Im(w_t)\in[-\frac{\eta}{2i},\frac{\eta}{2i}]$, as shown by the asterisks in subgraphs (b) and (c) of Fig.\ref{is-odd-fig}.
Thus, the ground state energies and surface energies are also expressed by Eqs.(\ref{Eigxy-odd}) and (\ref{Eisxy-odd}), respectively.

\begin{figure}[htbp]
\centering
\subfigure[Even $N$ case]{
\includegraphics[width=6cm]{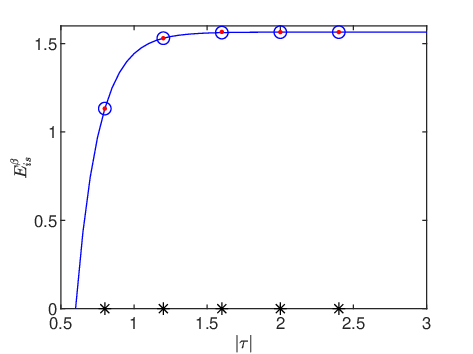}
}
\subfigure[Odd $N$ case]{
\includegraphics[width=6cm]{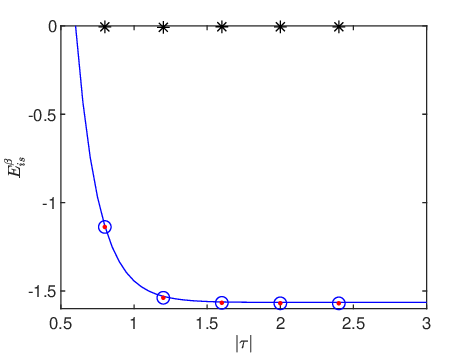}
}
\caption{
The surface energies $E^{\beta}_{is}$ with $\beta=x, y, z$ versus the model parameter $\tau$ for $\eta=0.6i$.
The blue lines denote the analytic results obtained from Eqs.(\ref{Eisxy-even}) and (\ref{Eisxy-odd}).
The red dots, blue circles and black asterisks denote the DMRG results for $E^{x}_{is}$, $E^{y}_{is}$ and $E^{z}_{is}$, respectively. (a) DMRG results with $N=150$, (b) DMRG results with $N=151$.
}\label{Esifig}
\end{figure}

Now, we check the analytic results (\ref{Eisz}), (\ref{Eisxy-even}) and (\ref{Eisxy-odd}) for the surface energies by using the numerical DMRG method, and the results are shown in Fig.\ref{Esifig}.
We see that the analytic results coincide with the DMRG results very well.
Furthermore, in the trigonometric limit $\tau\rightarrow i\infty$, the system reduces to the anisotropic XXZ model, which exhibits a gap.
These results are also consistent with the results of XXZ model in the gap region.

Next, we consider the first excited state.
For the boundaries $\beta=0, z$ with even $N$, there exist two real discrete zeros which have non-zero contributions to the energy, as shown by the circles in subgraphs (a) and (d) of Fig.(\ref{is-even-fig}).
From Eq.(\ref{Eroot-i}), we know that these two discrete zeros should be $\pm \frac{1}{2}$ in the thermodynamic limit.
Thus the energies $E^{\beta}_{ie}$ at the first excited states are
\begin{equation}
E^{0}_{ie}=E^{z}_{ie}=e_{i}N + 2E^{w}_{i}(\frac{1}{2} ). \label{Eie0z-small1}
\end{equation}
Then the related excitation energies are
\begin{equation}\label{gap-is}
 \Delta E^0_{i} =  \Delta E^z_{i}= 2E^{w}_{i}(\frac{1}{2}).
\end{equation}

For the boundaries $\beta=0, z$ with odd $N$, there exists one real discrete zero which has non-zero contributions to the energy, as shown by the circles in subgraphs (a) and (d) of Fig.\ref{is-odd-fig}.
The discrete zero should be $\frac{1}{2}$ in the thermodynamic limit.
Consequently, the energies at the first excited state read
\begin{equation}
E^{0}_{ie}= E^{z}_{ie}=e_{i}N + E^{w}_{i}(\frac{1}{2}). \label{Eie0z12}
\end{equation}
Therefore, the system has the continuous energy spectrum and
\begin{equation}\label{nogap-ib2}
\Delta E^0_{i}=0, \quad \Delta E^{z}_{i} =0.
\end{equation}

We consider the boundary conditions with $\beta=x, y$. If $N$ is even, there exists one real discrete zero, as shown by the circles in subgraphs (b) and (c) of Fig.\ref{is-even-fig}.
The discrete zero should be $\frac{1}{2}$ in the thermodynamic limit. Thus the energies $E^{x}_{ie}$ and $E^{y}_{ie}$ are
also expressed by (\ref{Eie0z12}), and the related excitation energies $\Delta E^{y}_{r}$ and $\Delta E^{z}_{r}$ are zero.
While if $N$ is odd, there exist two real discrete zeros which contribute the non-zero values to the energy, as shown in subgraphs (b) and (c) of Fig.\ref{is-odd-fig}. From Eq.(\ref{Eroot-i}).
These two discrete zeros should be $\pm \frac{1}{2}$ in the thermodynamic limit. The related energies $E^{x}_{ie}=E^{y}_{ie}$ are expressed as
(\ref{Eie0z-small1}) and the excitation energies $\Delta E^x_i=\Delta E^y_i$ are also given by Eq.(\ref{gap-is}).

In summary, the excitation energy of the system depend on the boundary conditions and the parity of $N$.
For the boundary conditions $\{\beta=0, z\}$ with even $N$, and boundary conditions $\{\beta=x, y\}$ with odd $N$,
the excitation energy is finite. The values of excitation energy are
\begin{equation}
  \Delta E_{i} =\left\{
              \begin{array}{cc}
               2E^{w}_{i}(\frac{1}{2}  )  ,& \Im(\eta) \in(0,\frac{\tau}{2i}], \\
                E^{w}_{i}(\frac{1}{2} +(\eta - \frac{\tau}{2}) ) + E^{w}_{i}(\frac{1}{2} -(\eta - \frac{\tau}{2}) ) ,&  \Im(\eta) \in(\frac{\tau}{2i},\frac{\tau}{i}).  \\
              \end{array}
            \right. \label{gapi}
\end{equation}
We shall note that the results at the point of $\eta=\frac{\tau}{2}$ can be obtained through the analytic extension of Eq.(\ref{gapi}).

\begin{figure}[!htp]
    \centering
\subfigure[]{
\includegraphics[width=6cm]{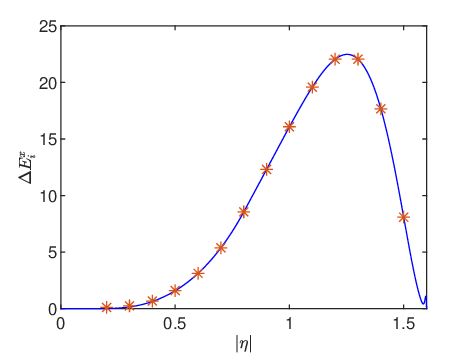}
}
\subfigure[]{
\includegraphics[width=6cm]{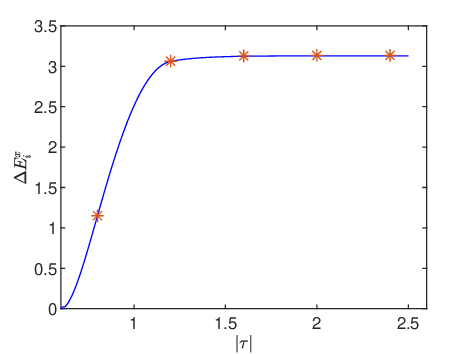}
}
\caption{
(a) The excitation energy $\Delta E^{x}_{i}$ versus the crossing parameter $\eta$ for $\tau=1.6 i$.
(b) The excitation energy $\Delta E^{x}_{i}$ versus $\tau$ for $\eta=0.6i$.
The solid lines are the analytic results obtained from Eq.(\ref{gapi}) and the red asterisks indicate the DMRG data with $N=311$.
}\label{fgapi}
\end{figure}

Now, we check the correctness of analytic results (\ref{gapi}) by the DMRG method. Without losing generality, we choose the excitation energy $\Delta E^{x}_{i}$ during the numerical calculation.
The results are shown in Fig.\ref{fgapi}. From it, we see that the analytic results coincide with the numerical ones very well.
Additionally, as shown in Fig.\ref{fgapi} (b), the results with the limit $|\tau|\rightarrow \infty$ give the excitation energy of the XXZ spin chain in the gapped region.

\section{Conclusions}

In this paper, we investigated the thermodynamic limits of the XYZ spin chain with the periodic or the twisted boundary conditions. Based on the new scheme of parameterization of eigenvalue of the transfer matrix by the zeros instead of Bethe roots,
we obtain analytically the ground state energy density, surface energy and the excitation energy. We find that these physical quantities depend on the parity of the system-size $N$ and the boundary conditions, due to
the different easy-axis and the anisotropic couplings.

For the real $\eta\in(0,1)$, the spins in the Neel state and in the low-lying states prefer to be arranged along the $x$-direction because it is the easy-axis. Thus the surface energy induced by the twisted boundary condition along the $x$-direction
has the same parity dependence as that for the ground state energy of the periodic system.
This dependence is broken by the twisted boundaries along the $y$- and $z$-directions. For the pure imaginary $\Im(\eta) \in (0,\frac{\tau}{i})$, the easy-axis is the $z$-direction.
The spins in the Neel state and in the low-lying states prefer to be arranged along the $z$-direction.
Thus the energies in the thermodynamic limit induced by the twisted boundary condition along the $z$-direction have the same parity dependence as
that for the periodic case. This dependence is broken by the twisted boundaries along the $x$- and $y$-directions.

\section*{Acknowledgments}

We acknowledge the financial support from National Key R$\&$D Program of China (Grant No.2021Y FA1402104),
National Natural Science Foundation of China (Grant Nos. 12434006, 12205235, 12247179, 12247103, 12105003, 12074410 and 11934015),  Shaanxi Fundamental Science Research Project for Mathematics and Physics (Grant Nos. 22JSZ005), and the Strategic Priority Research Program of the Chinese Academy of Sciences (Grant No. XDB33000000).

\section*{Appendix: Elliptic functions and identities}
\label{appA}
\setcounter{equation}{0}
\renewcommand{\theequation}{A.\arabic{equation}}

During the derivation in this paper, we have used following properties of the elliptic functions.
The $\sigma$-function satisfies the Riemann-identity
\begin{eqnarray}
  && \sigma(u+x) \sigma(u-x)\sigma(v+y)\sigma(v-y) - \sigma(u+y) \sigma(u-y)\sigma(v+x)\sigma(v-x)  \nonumber\\
  &=& \sigma(u+v) \sigma(u-v)\sigma(x+y)\sigma(x-y),
\end{eqnarray}
and the ones
\begin{eqnarray}
  && \sigma(2u)= \frac{2 \sigma(u)\sigma(u+\frac{1}{2}) \sigma(u+\frac{\tau}{2}) \sigma(u-\frac{1+\tau}{2})     }{ \sigma(\frac{1}{2}) \sigma(\frac{\tau}{2}) \sigma(-\frac{1+\tau}{2}) },  \label{doubleang}\\
  && \frac{\sigma(u)}{\sigma(\frac{\tau}{2})} = \frac{ \theta \left[
           \begin{array}{c}
             0 \\
             \frac{1}{2} \\
           \end{array}
         \right](u,2\tau) \theta \left[
           \begin{array}{c}
             \frac{1}{2} \\
             \frac{1}{2} \\
           \end{array}
         \right](u,2\tau) }{ \theta \left[
           \begin{array}{c}
             0 \\
             \frac{1}{2} \\
           \end{array}
         \right](\frac{\tau}{2},2\tau) \theta \left[
           \begin{array}{c}
             \frac{1}{2} \\
             \frac{1}{2} \\
           \end{array}
         \right](\frac{\tau}{2},2\tau) }, \\
  && \theta \left[
           \begin{array}{c}
             \frac{1}{2} \\
             \frac{1}{2} \\
           \end{array}
         \right](2u,2\tau) =  \theta \left[
           \begin{array}{c}
             \frac{1}{2} \\
             \frac{1}{2} \\
           \end{array}
         \right](\tau,2\tau) \times \frac{\sigma(u)\sigma(u+\frac{1}{2}) }{ \sigma(\frac{\tau}{2})\sigma(\frac{\tau}{2}+\frac{1}{2}) } , \\
  && \theta \left[
           \begin{array}{c}
             0 \\
             \frac{1}{2} \\
           \end{array}
         \right](2u,2\tau) =  \theta \left[
           \begin{array}{c}
             0 \\
             \frac{1}{2} \\
           \end{array}
         \right](0,2\tau) \times \frac{\sigma(u-\frac{\tau}{2})\sigma(u+\frac{1}{2} + \frac{\tau}{2}) }{ \sigma(-\frac{\tau}{2})\sigma(\frac{\tau}{2}+\frac{1}{2}) }.
\end{eqnarray}


\providecommand{\href}[2]{#2}\begingroup\raggedright\endgroup

\end{document}